\newcommand{\xmm}{XMM-Newton}

\newcommand{\dd}{{\rm d}}
\newcommand{\lln}{{\rm ln}}
\newcommand{\lya}{${\rm Ly}\alpha$}

\newcommand{\ann}[1]{$^{\rm #1}$}
\newcommand{\celsius}{$\,^\circ{\rm C}$}

\newcommand{\dg}{$^\circ$}
\documentclass{aa}
\usepackage{natbib}
\usepackage{graphicx}
\usepackage{txfonts}
\usepackage{supertabular}
\usepackage{amsmath,mathtools}
\usepackage[]{changebar}
\usepackage[dvips]{color}

\begin{document}
   \title{Calibration and in-orbit performance of the reflection grating spectrometer onboard \xmm}

   \author{C.P.~de~Vries\inst{1}
          \and
	  J.W.~den~Herder\inst{1}
	  \and
	  C.~Gabriel\inst{2}
	  \and
	  R.~Gonzalez-Riestra\inst{2}
	  \and
	  A.~Ibarra\inst{2}
	  \and
	  J.S.~Kaastra\inst{1}
	  \and
	  A.M.T.~Pollock\inst{2}
	  \and
	  A.J.J.~Raassen\inst{1,4}
	  \and
	  F.B.S.~Paerels\inst{3}
          }

   \offprints{C.P.de.Vries@sron.nl}

   \institute{
     SRON Netherlands Institute for Space Research,
     Sorbonnelaan 2, 3584 CA Utrecht, The Netherlands
     \and
     European Space Agency, ESAC, Apartado 50727, 28080, Madrid, Spain
     \and
     Columbia Astrophysics Laboratory, 550 West 120th Street, 
     New York, NY 10027, USA
     \and
     Astronomical Institute 'Anton Pannekoek', Science Park 904, 1098 XH Amsterdam, 
     University of Amsterdam, The Netherlands
   }

   \date{Received 24 Februari 2014 /  Accepted 16 October 2014}

 
  \abstract
   {
   \xmm{} was launched on 10 December 1999 and has been operational since early 2000.
   One of the instruments onboard \xmm{} is the reflection grating spectrometer (RGS). 
   Two identical RGS instruments are available, with each RGS combining a reflection grating assembly (RGA) and a
   camera with CCDs to record the spectra. 
   }
   {
   We describe the calibration and in-orbit performance of the RGS instrument. By
   combining the preflight calibration with appropriate inflight calibration data
   including the changes in detector performance over time, we aim at profound
   knowledge about the accuracy in the calibration. This will be crucial for any
   correct scientific interpretation of spectral features 
   for a wide variety of objects. 
   }
   {
   Ground calibrations alone are not able to fully characterize the instrument.
   Dedicated inflight measurements and constant monitoring are essential for a full understanding
   of the instrument and the variations of the instrument response over time. Physical
   models of the instrument are tuned to agree with calibration measurements and are the basis
   from which the actual instrument response can be interpolated over the full parameter space.  
   }
   {
   Uncertainties in the instrument response have been reduced to 
   $< 10 \%$ for the effective area and $<$ 6 m\AA{} for the wavelength
   scale (in the range from 8 \AA{} to 34 \AA). The remaining
   systematic uncertainty in the detection of weak absorption
   features has been estimated to be 1.5\%.
   }
   { 
   Based on a large set of inflight calibration data and comparison
   with other instruments onboard \xmm{}, the calibration accuracy of the
   RGS instrument has been improved considerably over the 
   preflight calibrations.
   }


\titlerunning{Calibration and In-orbit Performance of the RGS}

   \maketitle
%

\section{Introduction}

Accurate understanding of the instrument response is a prerequisite
for correctly interpreting the observational data from \xmm. This understanding 
requires a profound knowledge of the limitations of the preflight
calibrations and the models applied to calculate the response in orbit. 
In addition, changes in the calibration due to the operational conditions 
in space (radiation damage, contamination, etc.)
should be understood well and be taken into account.
Apart from statistical fluctuations in the data, important scientific results also
depend on understanding the systematic errors of the instrument.
With the simultaneous operations of two RGS instruments, we have an excellent
tool for studying systematic errors. If the instrument
model is correct, the two RGS instruments should produce consistent results
within the statistical uncertainty. By using high signal-to-noise observations we can test this in detail. 


The prime goal of the RGS instrument is to measure the emission lines and absorption features of specific 
transitions in highly ionized plasmas. In addition to a reasonable effective area, this requires especially
good spectral resolving power (R $>$ 250). Clearly, identifying
and quantifying weak absorption or emission features is very challenging. 
For example, weak absorption features due to the warm hot intergalactic medium (WHIM) in the direction of the bright blazar
Mrk 421 have been reported and disputed \citep{Nicastro2005,Kaastra2006,Rasmussen2007}. Because these detections
have a typical significance between 3, and 5 $\sigma$, an accurate
calibration of the instrument response is essential. Also studying the
velocity broadening of lines critically depends on understanding the 
response (see section \ref{sec:lsf}). Knowing the time evolution of the response is important as well. 
The time-dependent change in the spectrum of 
the X-ray emission from the isolated neutron star RXJ 0720.4-3125 \citep{devries2004,Haberl2006}
could only be detected using detailed knowledge about the
stability of the RGS response. 

\vskip 0.1 cm

In this paper we give a quantitative assessment of the calibration of the RGS instrument onboard
{\it XMM-Newton}. In section~\ref{sec:instrument} we present the main characteristics of the
instrument. In section~\ref{sec:calibration_basics} we describe the calibration method and
data processing steps applied in the current data processing system. 
This is followed by the main part of this
paper where we describe the preflight and inflight calibration for the various
key instrumental characteristics (effective area, line spread function, wavelength calibration, and background).
In the last sections (\ref{sec:lsf},\ref{sec:wavcal},\ref{sec:effarea}) we summarize the current status of
the calibrations and give an an overview
(section~\ref{sec:operations}) of relevant operational information. 

Since this paper aims to give a total overview of RGS performance and calibration, it is in part also a review of
papers and calibration and operational documents published at various places elsewhere, in particular at the \xmm{}
science center at ESAC. Besides new information and data also previously published data are shown, when appropriate,
to present a complete picture.   

The latest calibration data are used by the standard data processing package for \xmm{}, the 
science analysis system (SAS), which is maintained and can be downloaded from the \xmm{} Science
Operations Center at ESAC. 
In particular the "Calibration access and data handbook" offers many details
on the calibrations and algorithms used.

\section{Instrument}

\label{sec:instrument}
\begin{figure} 
\resizebox{\hsize}{!}{\includegraphics{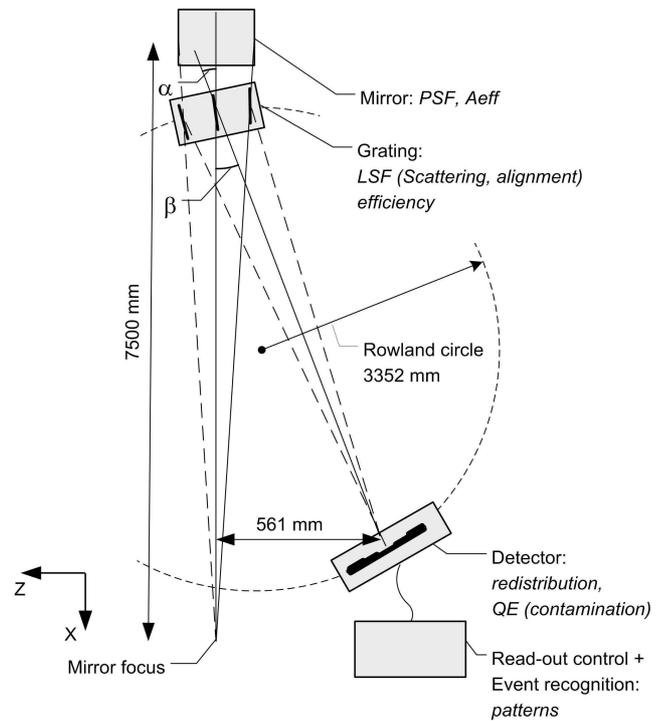}}
\caption{Schematics of the RGS configuration (not to scale)}
\label{fig:schematics}
\end{figure}

Both RGS instruments (RGS1 and RGS2) consist of a set of reflection gratings placed in the converging beam of 
the telescope combined with a camera in the spectroscopic focus. Detailed information about the instrument is given
in \cite{herder} but some key features of the instrument, that are important to understand
the accuracy of the calibrations are repeated here. The instrument is shown in Fig.~\ref{fig:schematics} where 
the main components are given including the related 
calibration dependencies. The mirror response is characterized by its effective area ($A_{\rm eff}$) and its point spread 
function (PSF). In the converging beam the  reflection gratings are positioned on the Rowland circle. The response
of the gratings can be described by their reflection efficiency, the scattering and the grating-to-grating alignment. 
The reflected X-rays are recorded on a photon-by-photon basis in the CCD detector where the position corresponds to 
the wavelength (energy) and the energy resolution of the CCDs is used to separate orders. The performance of these detectors will 
vary over the mission lifetime due to radiation damage (e.g. change in the CCD redistribution function) and due to 
contamination on the detectors (change in quantum efficiency). Finally X-ray event patterns on the CCDs are
recognized by the read-out unit. 
The european photon imaging camera (EPIC-MOS) is located in the focus of the mirrors.

\vskip 0.3 truecm

The central position of the X-rays on the detector is given by the dispersion equation through:

\begin{equation}
 m \lambda = d ~ ( \cos ~ \beta ~ - ~ \cos ~ \alpha) 
\end{equation}

where $m$ is the spectral order (--1, --2), $d$ is the groove spacing, and $ \alpha $ and 
$\beta$ are the angles of the incident and dispersed rays measured from the grating plane.
First and second order spectra are overlapping on the detectors and using the energy resolution of the 
detector they can be easily separated. 
For extended sources the wavelength resolution degrades to

\begin{equation}
\Delta \lambda = 0.124 ~ \times ~ \theta ~ / ~ m  
\end{equation}

where $\theta$ is the source extent in arc min and $\Delta \lambda$ in \AA. The instrument is optimized at 15~\AA\ around the
Fe-L lines.
Key parameters of the design are listed in Table \ref{table:design}.

\vskip 0.1 cm

\begin{table}
\caption{Key design parameters}
\label{table:design}
\centerline{
\begin{tabular}{rrp{4cm}}
\hline
parameter             	& value                         & comment \\
\hline
$\lambda\rm_{blaze}$    & 15 \AA        	& first order blaze wavelength \\
$\delta$        	&0.6989\dg       	&blaze angle of facets\\
$\gamma$        	&2.2751\dg        	&graze angle on facets \\
$\alpha$        	& 1.5762\dg		&angle of incidence \\
$\beta\rm_{blaze}$      & 2.9739\dg 		&diffraction angle for $\lambda_{\rm blaze}$ \\
$1/d$             	& 645.6~lines/mm 	&central groove density\\
CCDs            	& 9                     & back illuminated, $\sim 30 \mu$m thick \\
pixels          	& 27 $\times$ 27 $\mu$m$^2$     & pixel size; standard $3\times3$ on chip binning  \\
CCD size        	& 1024 $\times$ 384 		& image area, identical storage area \\
T               	& $-80$\ to \ -110\celsius  \\
\hline
$\lambda$-range 	& 6.0 - 38 \AA 		& \\
$A_{\rm eff}$		& $140~{\rm cm}^2 $ 		& for two units at 15 \AA \\
resolving power		& 100 - 500		& $\approx 250$ at 15 \AA \\
$\Delta \lambda $ 	& 8~m\AA 		& wavelength accuracy \\
\hline\noalign{\smallskip}
\end{tabular}
}
\end{table}

Each instrument consists of four different units: the {\it reflection grating assembly (RGA)}
which is an assembly of accurately positioned  gratings; the {\it RGS focal plane camera (RFC)}
which is an array of 9 CCDs in which the reflected photons are detected; the RGS {\it analog electronics (AE)}
which contains part of the data processing chain and controls the read-out sequence of the CCDs and the
RGS {\it digital electronics (DE) } which controls the instrument and performs event pattern recognition. 

\vskip 0.1 cm

{\it Reflection gratings (RGA): } 182/181 in-plane reflection gratings for RGS1/2 are accurately placed in the
exit beam of the X-ray
telescopes. Each grating is 200 mm long by 100 mm wide with a gold reflecting surface and a ruggedized backside
with 5 supporting ribs to control the surface errors. Each grating is mounted at its four corners.
To correct for the beam convergence, the line density of the gratings varies with position $x$ on the grating in such a way
that line density is 645.6 lines/mm at grating center, while being 665.3 lines/mm at $x=+100$~mm and
626.8 lines/mm at $x=-100$~mm. 
The gratings are mounted in a Rowland circle configuration, on a support structure made
of Be because of its mass and low thermal expansion coefficient.
To maintain the proper focusing of all gratings the temperature gradient over the
grating assembly should be less than 1$^\circ$C.

\vskip 0.1 truecm

The X-rays are recorded by an array of 9 CCDs in the {\it RGS focal plane camera (RFC)}. 
This camera is passively cooled to reduce the dark current to an 
acceptable level ($\leq 0.06 ~{\rm e}^-{\rm /pixel/s}$).
To have a good QE ($\geq$ 80\%) at low energies, the CCDs are
illuminated from the backside, avoiding the absorption that would
otherwise be caused by the gate structure at the front side of the CCDs.
To reduce the sensitivity for optical load, the CCDs
are coated with a thin aluminum layer with a thickness of 45 nm, 60 nm
and 70 nm depending on the relative position of the CCD with respect to the
mirror focus. To isolate the Al layer from
the CCD itself an insulating ${\rm MgF}_2$ layer has been applied. These
components produce absorption edges in the response of the instrument.
Subsequently the X-rays are absorbed in the high-epitaxial Si ($\geq 10 ~\Omega{\rm m}$) \citep{jansen1989}. 
For each absorbed photon the
number of electrons is proportional to the energy of the incoming photon.
To avoid unnecessary diffusion of the produced
electron cloud in the Si prior to charge cloud collection and read out
near the gate structure, the CCDs are thin ($30 \mu m$). This implies
that the QE around the Si-edge is less than 100 \%. In addition, some charge will be lost for
events absorbed near the backside of the CCDs. This causes a partial event tail in the CCD redistribution function. 

All CCDs have an image and a storage section and are read out in frame transfer mode. The 
clock pattern for this read-out is programmable and it is possible to perform on-chip binning, to
read out only parts of the CCD, or to read out the full CCD through one or both of its two output nodes.
During standard operations the charge of
$ 3 \times 3 $ physical CCD pixels of $ 27 \mu {\rm m}~ \times ~27 \mu {\rm m}$ is added
on the CCD (these $ 3 \times 3 $ pixels are referred to as a bin).
Still, the charge generated by a single X-ray event can be split
over more than one bin causing so-called split events. To ensure the direct
relation between charge and photon energy, pile-up (two events in the same bin
or in neighboring bins 
during a read-out) should be avoided. The typical read-out time per CCD, with the two CCD halves read out simultaneously
through both output nodes, is $\approx 0.6$ s,
resulting in a readout time of 4.6 s for the array of 8 operational CCDs. 
Except for very bright sources such as e.g. Sco X-1 or the Crab 
the pile-up is very small (normally well below 1-2 \%).

In addition to X-rays dispersed by the gratings, the cameras also register X-rays from four onboard calibration
sources located in the thermal/radiation shield just around the detector. $\alpha$ particles from a Cm244
source illuminate an Al target generating Al-K line emission in two sources. 
For the two other sources a Teflon target has been selected, generating F-K$\alpha$ emission.
The generated X-rays do not overlap with the image of a point source and their energies do not
interfere with the 1st and 2nd order dispersed source spectrum. 

\vskip 0.1cm
The {\it analog electronics (AE) } controls and selects the CCD to be processed and measures
the amount of charge in the output node of the CCD. It
amplifies the signal and adds an electronic offset, before passing the signal to the ADC of the digital
electronics.   

\vskip 0.1cm

The {\it digital electronics (DE)} processes the data. Bins
with only the dark current and electronic noise (lower threshold) and bins with a too high
charge due to a charged particle hitting the detector (upper threshold) are immediately rejected. 
The remaining bins are searched for
acceptable event patterns (combination of adjacent bins within a $ 2 \times 2 $ bin
region which are characteristic for a true X-ray event). Patterns larger than
a $2 \times 2$ bin region are rejected. The DE also stores a list of uploaded hot pixels and
columns, used for rejection of these items. 
The relative loss of CCD area due to interactions of charged particles is computed for each CCD readout frame. 
Later, on ground, the effective exposure time for each CCD readout frame is corrected 
in correspondence with this relative loss of CCD area and for the known hot pixels and columns.

In addition to this routine data processing, full frames
are stored at a very low rate of one every 15 minutes and using spare telemetry capacity these full
frames are transferred to ground. This provides diagnostic data to 
monitor the CCD performance in full detail.
A cold redundant digital electronic unit is present. 

\vskip 0.1 cm

\begin{figure*}
\resizebox{7in}{!}{\includegraphics[clip]{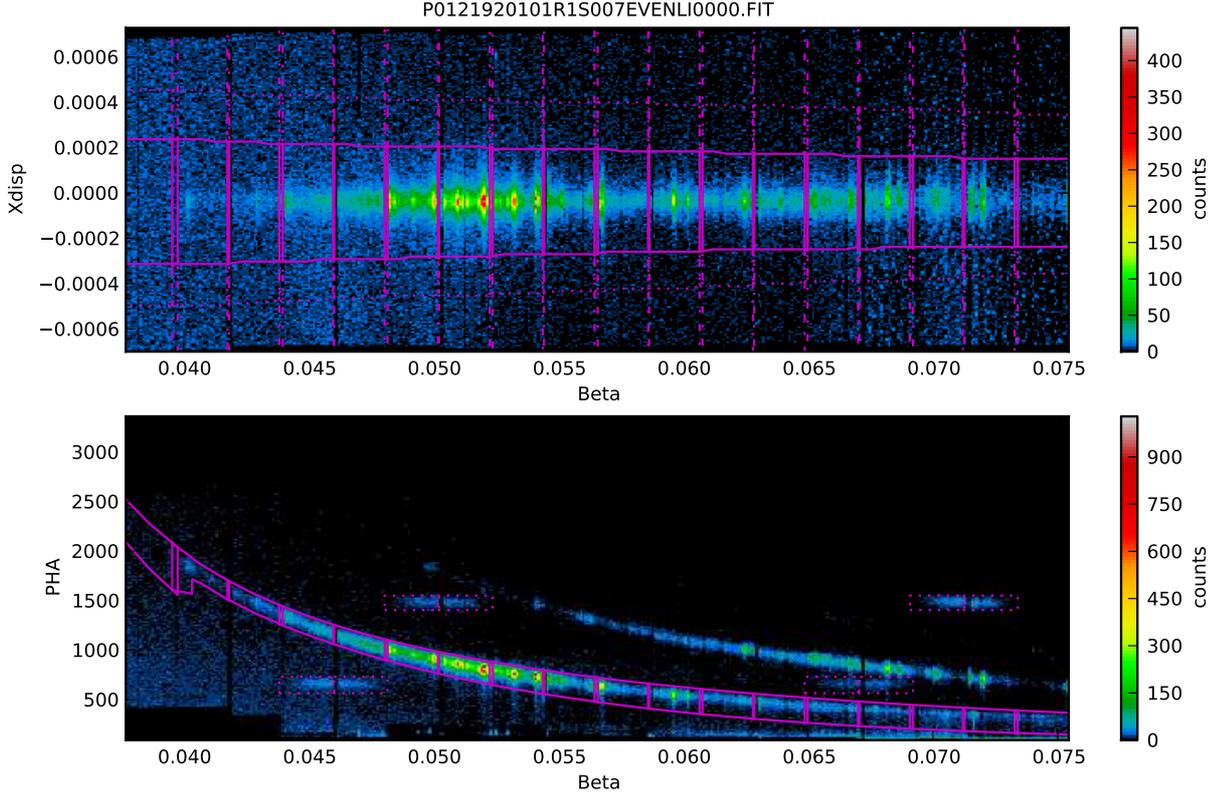}}
\caption{Capella observation (id: 0121500201001). The color scale represents a logarithmic intensity scale.
The dispersion (Beta) axis runs horizontally and increases to the right. All angles (Beta, Xdisp) are in radians.
In the top panel the cross dispersion (Xdisp)
is along the vertical axis whereas in the bottom panel this is replaced
by the CCD recorded 
photon energy, or PHA, in units of eV.
In the bottom panel the low level thresholds
are clearly visible. Standard data selections for first order are indicated by the solid purple curves.}
\label{fig:data_space}
\end{figure*}

The main observable parameters are the position along the dispersion axis (the photon wavelength), the position
in the cross dispersion direction and the photon energy as recorded by the pulse-height 
in the detector (for order separation).
In Fig.~\ref{fig:data_space} these data are shown for a point source. In addition to the X-rays from the celestial source,
the four on-board calibration sources with different energies (F-K$_\alpha$ at 0.677 keV
and Al-K$_\alpha$ at 1.487 keV) are clearly visible.

\section{Calibration basics}

\label{sec:calibration_basics}

\subsection{Calibration elements}

The response of an instrument describes the conversion from photons incident on the instrument to the
data actually recorded by the instrument. For a spectroscopic instrument like RGS the response key elements are
effective area, wavelength scale, and linespread function. We define these elements as follows:

{\it Effective area:} This quantity is a function of energy ($A_{\rm eff}(E)$), and is defined as the 
ratio between the observed source count rate (${\rm counts\,s}^{-1}$) and the incident photon
flux (${\rm photons\,s}^{-1}{\rm cm}^{-2}$) For RGS the effective area can be decomposed into the product
of the following different components:

\begin{itemize}
\item{Mirror area $\times$ mirror efficiency}
\item{Grating efficiency}
\item{CCD quantum efficiency}
\item{Exposure fraction}
\item{Data selections}
\end{itemize}


The different components depend on a multitude of parameters.
The mirror geometrical area is the front surface area of the total
instrument. The CCD quantum efficiency (QE) includes all losses in the absorbing and dead layers on the CCD,
charge transfer efficiency (CTE), onboard signal thresholds and processing etc., and depends on the actual CCD chip and
location on the chip. The exposure fraction is relative to the real observing time. Apart from dead times caused
by onboard processing limitations and cosmic ray hits it also includes the effects of instrument pointing variations.
These variations move photons to different pixels on chip locations with different QE or dead pixels. 
Subsequent data selections also affect the number of recorded photons.

{\it Linespread function (LSF):} This is the probability distribution of the location where photons of given energy land
on the detector, ${\rm LSF}(E,pointing,..)$. This depends on the quality
and internal alignment of the optics, both in terms of optical surface accuracy and scattering contamination.
Pointing reconstruction accuracy and data selections also contribute.

{\it Wavelength scale:} This function is defined as the relation between the energy of the incoming photon and the most 
likely location, which is the center of the linespread function, where it will end up on the detector, $W(E)$. 
This depends on pointing, grating parameters and the internal alignment of the different instrument components.

Table~\ref{table:dependency} gives an overview of the different dependencies. In the next sections
these components are treated in more detail.

All response elements may change over time, but during the operational period a few major adjustments
caused sudden changes to specific components. Further cooling of the detector, change of CCD readout mode and adjustment of
CCD bias voltage had major impact on the response. Details of these instrumental operations are described in
section~\ref{sec:operations} and listed in table~\ref{tab:events}.
 
\begin{table*}
\caption{Calibration dependency}
\label{table:dependency}
\begin{center}
\begin{tabular}{rl}
\hline
parameter	&	dependency \\
\hline
line spread function	& mirror PSF \\
		& gratings alignment \\
		& gratings scattering \\
		& particle contamination of the optics \\
		& \hspace{1cm} (causes additional scattering) \\
	        & pointing reconstruction accuracy \\
\hline
wavelength scale & grating dispersion parameters \\
		& mirror -- grating assembly alignment \\
		& focal plane camera position \\
		& CCD positions in the camera \\
		& stability of optical bench \\
		& pointing reconstruction accuracy \\
\hline
effective area	& mirror effective area \\
		& geometrical overlap gratings/mirrors \\
		& grating efficiency 	\\
                & CCD quantum efficiency \\
		& \quad \quad \quad CCD molecular contamination \\
		& \quad \quad \quad CCD 'absorption' layers (filters) \\
		& \quad \quad \quad CCD thickness \\
		& \quad \quad \quad event pattern processing \\
	        & CCD redistribution \\
		& \quad \quad \quad read-out noise \\ 		
		& \quad \quad \quad CCD dark current \\
		& \quad \quad \quad charge transfer inefficiency \\
		& \quad \quad \quad X-ray event pattern \\
		& exposure \\
		& \quad \quad \quad satellite pointing \\
		& \quad \quad \quad CCD cosmetic defects \\
		& \quad \quad \quad particle background \\
		& data selections \\
\hline

\end{tabular}
\end{center}
\end{table*}

\subsection{Calibration method}

Accurate calibration is a long and tedious process. As a full mapping of the
response over all energies and all source positions is impossible, we combine models, constrained or verified by real
measurements. Subsequently this model was updated using in-orbit data.

The calibration starts with the
physical understanding and measurement of the response for different units including the
mirror, the detector (first as CCD, subsequently as integrated camera) and the reflection
grating assembly (first as grating plate, later as integrated RGA). 

Following the detailed calibration of the physical models, the instrument
response is compared with end-to-end performance
tests at the MPE Panter facility \citep{Freyberg2005,Brauninger2004} at specific energies and positions in the field of view. 
The major limitations of this approach are:

\begin{itemize}

\item{
 the measurements can be performed only for a limited
number of energies and angles; 
}

\item{no X-ray sources with narrow emission lines below $\sim 1$ keV exist;
}

\item{due to the finite source distance the first part of the mirror is not fully illuminated resulting in a
not fully representative measured mirror response;
}

\item{the alignment of the units cannot be tested on the ground as the correction for the
finite source position (120 m) requires a different relative alignment of the units than in orbit.
} 

\end{itemize}

\vskip 0.1 cm

To cope with these limitations for the ground calibration a dedicated calibration campaign in the first months in orbit 
was foreseen. With a number of astrophysical sources unknown calibration
parameters, such as the wavelength scale, could be fixed.

\vskip 0.1 cm

Following this early phase in-orbit calibration a set of 'routine' 
calibrations are performed with different aims:

\begin{itemize}

\item{ {\it Effective area stability}: The stability of the effective area is
checked by bi-annual measurements of two stable soft sources: the isolated neutron star RXJ1856-3754 and
the Vela Pulsar Wind Nebula. Two other sources turned out to be
less suited: the isolated neutron star RX J0720.4-3125 showed to be variable \citep{devries2004}, and
SNR 1E0102.2-7219 (or N132D) is an extended source. In extended sources, spatial and spectral information is mixed,
which means that if the source is not circularly symmetric, the extracted spectrum depends on the roll angle
of the satellite, and is therefore
less suitable for a regular verification of the effective area.
Measurements of a strong continuum source (BL Lacs PKS 2155-304 and Mrk 421) allow a check on
the stability of the oxygen edge, which can change over the mission lifetime due to the 
accumulation of ice on the cooled detector. 
}

\item{ {\it Wavelength scale}: Initially a set of coronal sources
were measured regularly. AB Dor once a
year, Capella and HR 1099 alternating also once a year. Recently 
this has been reduced to Capella once a year only
because the wavelength scale does not show long term trends.
}

\item{ {\it CCD response}: Dedicated measurements with Mrk 421 on and off-axis
allows the measurement of the  CTI in the detectors. For different positions
off-axis in the cross-dispersion direction, the number of parallel
transfers in the CCD varies, and this allows
an accurate calibration of the parallel CTI. The sides of two adjacent CCDs
are read out in opposite direction. For RGS2, after revolution 1408, all pixels are
read out trough one single node. Due to these serial transfers the pixels near
the CCD edges are hardly affected whereas the pixels in the center, with typically
512 or 1024 transfers, show a reduction of a few percent
in pulse height. This is observed as a saw tooth
shape of the recorded CCD pulse heights versus dispersion direction in the spectrum,
that disappears after calibrating the serial CTI.

In addition, the observations of Capella are used to verify the CCD electronic
readout gain, defined as the conversion factor from captured electric charge to pulse height readout units, 
and the serial CTI calibration of the CCDs. 

Furthermore, full frame CCD images, obtained during all observations, allow us to monitor the development
of artifacts due to radiation damage in the CCDs and the changes in the read-out noise and dark current
of the CCDs.
}

\item{ {\it Non-routine calibrations}: In addition, non-routine calibrations
are carried out to investigate certain aspects. A detailed measurement of the
Crab spectrum was carried out with the aim of calibrating the absolute
effective area \citep{Kaastra2009} and of Sco X-1 with the aim to fully
characterize the X-ray absorption edge structures present in the CCDs response \citep{devries2003}.
}

\end{itemize}

Some of these measurements were planned to coincide with the same 
target in the Chandra observatory observation program. Detailed comparison of our results
with those of Chandra is, however, outside the scope of the current
paper. Information on cross calibrations between high energy missions can e.g. be found in the
proceedings of the IACHEC meetings (http://web.mit.edu/iachec/)

\vskip 0.3 truecm

\begin{table*}
\caption{Current inflight calibration plan}
\label{table:inflight_plan}
\begin{center}
\begin{tabular}{rcl}
\hline
purpose	& frequency (/year) & measurement \\

\hline
$\lambda$ calibration 	& 1  	& coronal source: Capella \\
effective area		& 2	& continuum source: Mrk 421, PKS 2155-304 \ann{1} \\
			& 1	& Vela SNR \\
			& 1	& RXJ 1856--3754 \\
CCD response (gain/CTI)	& 0.5	& Mrk 421 on and off axis (CTI) \ann{1}\\
    			& 1   	& Capella (gain)\\
\hline

\end{tabular}
\vskip 0.2cm
\begin{minipage}{15cm}
\begin{itemize}
\setlength{\itemsep}{0pt}
\item[1)] During the period when Mrk 421 is not visible, it is replaced by PKS 2155-304
\end{itemize}
\end{minipage}
\end{center}
\end{table*}

\subsection{Calibration files}

The basic tool to analyze RGS spectra is the instrument response matrix (RSP).
The response matrix is the product of the mono-energetic X-ray redistribution (RMF), the effective area (ARF)
and exposure map.  Data channels are re-binned CCD columns
in the dispersion axis ($\beta$-angle) or in wavelength ($\lambda$) space. 
The exposure map contains for each data channel the relative exposure time of that channel
with respect to the exposure time of the observation. E.g. discarded bad CCD pixels or columns in the instrument, CCD dead
times due to lost telemetry packets are accounted for in the exposure map. In the total response matrix, all these
different elements of the calibration are combined. 

Over time the understanding of the instrument response will evolve, 
both in terms of additional or improved calibration data, and in terms of improved instrument models. 
In addition, the calibration of the instrument
is not static but varies over time. This is accommodated by the current calibration files (CCF).
These CCFs store the parametrized instrument response and are used by the instrument model. 
The CCFs are tagged with a time validity interval, usually spanning many observations, which allows
to have parameters which slowly change with time (gain, CTI, bad pixels). Changes can be accommodated
by updating the relevant CCF. Only occasionally the actual
data processing code needs to be updated to accommodate improvements in instrument models. 
The relevant CCFs are given in Table ~\ref{table:ccf_contents}. 
The actual CCFs, including their sequence number and validity start time are given in Table~\ref{table:CCF}. 
The CCFs were designed with flexibility in mind; some CCFs contain elements which 
later turn out not to be used.

Main steps in the RGS data calibration processing are as follows:

\begin{itemize}
\item{{\it Event processing.} From each recorded event, the CCD noise level is subtracted on a pixel by pixel basis, as derived from
a three orbit average computed from CCD images, and the recorded pulse level (PHA) is converted to eV using the gain
and CTI calibration. Warm/hot pixels are detected, and corresponding events discarded.}
\item{{\it Wavelength.} Using the spacecraft pointing information and instrument geometry, the conversion from
detector coordinates to $\beta$/wavelength space is computed.}
\item{{\it Spectra.} Using the given spatial, PHA, and time selections within the observation, source and background
spectra are computed in detector channel (either $\lambda$ or $\beta$ space).}
\item{{\it Response.} Using the appropriate calibrations for the time of observation, the response matrix is computed.}  
\end{itemize}

Instrumental background is usually obtained from off-source regions in cross dispersion on the detectors, using identical PHA
selections as for the source. 
This works very well for sources
which illuminate only a fraction of the RGS cross-dispersion field of view (4.9 arcminutes),
but cannot be used for extended sources which cover the entire detector.
For this reason a separate set of calibration files is maintained which provide backgrounds derived from
empty fields. These backgrounds are mainly generated by the fluctuating soft proton flux and cover a variety of different
flux levels. 
The background in this case is constructed by selecting the appropriate background calibration files, based
on the soft proton level indicators constructed from CCD9 count rates 
(see also section \ref{sec:softp}).

\begin{table}
\caption{CCF files, applicable to RGS}
\label{table:ccf_contents}
\begin{center}
\begin{tabular}{lp{5cm}}
\hline
CCF	&	contents \\
\hline
ADUConv			& gain (ADC to eV conversion) \\
			& average noise level (not used) \\
CTI			& parallel and serial CTI correction \\
BadPix			& bad pixels, segments and columns \\
			& h = uploaded, H = on ground only \\
CoolPix			& additional hot/cool pixels \\
Quantum Efficiency 	& CCD filter layers and Si thickness \\
			& RGA efficiencies and vignetting$^1$  \\
EXAFS			& ${\rm MgF}_2$ edge and O edge \\
EffAreaCor 		& contamination correction \\
			& correction for large scale variation\\
ReDist			& CCD PHA redistribution parameters \\
LineSpreadFunc 		& RGA figure/alignment errors and scattering parameters \\
CrossPSF		& source profile in cross dispersion \\
LinCoord		& relevant alignment parameters affecting wavelength scale\\
TemplateBackground	& spectra for pre-defined background  \\
			& intensities for data selections \\
XRT XAreaEff 		& mirror effective area and vignetting$^2$ \\
XRT XPSF 		& mirror RGS point source PSF profile, used for LSF computation \\
XMM BoreSight		& instrument alignment parameters \\
MiscData		& various additional parameters\\

HKParmInt		& RGS housekeeping parameter limits \\
SAACorr			& coefficients for the sun angle wavelength correction \\
ClockPatterns		& CCD clock pattern parameters \\
XMM ABSCoef		& Henke absorption coefficients for certain elements\\

\hline
\end{tabular}
\end{center}

\par 1) The geometrical effect that for photons leaving a grating at larger angles
the backside of the neighbouring grating may block part of the X-rays (shadow effect).
This depends on spacecraft pointing and energy of the photons. 
\par 2) The geometrical effect that for off-axis angles, the mirror shells may partially block
each other. Blocking is different for different parts of the mirror, which means there
will be some energy dependence, since reflection efficiency differs depending on photon energy and reflection
angle.

\end{table}

\section{Mirror calibration}

\begin{figure}
\resizebox{\hsize}{!}{\includegraphics [angle=90] {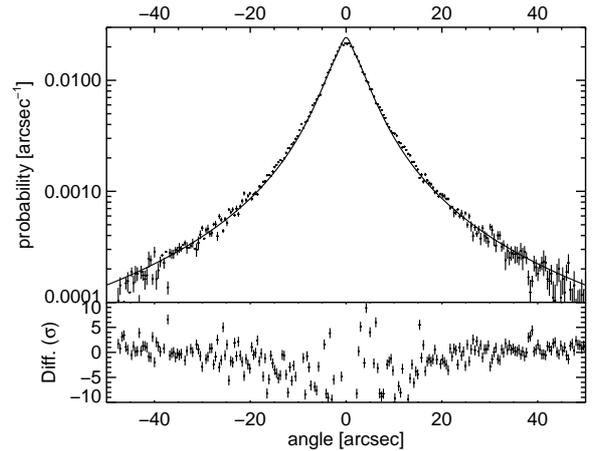}}
\caption{Measured and modeled mirror response (projected on the dispersion axis). Fit residues (lower plot) are
given in units of the statistical error.}
\label{fig:mirror}
\end{figure}

The two main functions describing the mirror response are
(a) the Point Spread Function and (b) the effective area. Of course they depend on the offset from the optical axis.
In Fig. \ref{fig:mirror} we show the measured mirror PSF, projected on the RGS
dispersion axis, and the mirror model used for RGS. 
The model consists of a combination of a Gaussian and a Lorentzian function, which are parametrized in the CCF. 

 
Although systematic effects dominate there is a good match,
in general within 2--4 $\sigma$ for the wings,
between
the model and the data with the exception of an overprediction towards the
core of the distribution. This can be explained by some pile-up in the core of the measured distribution. 
However, the angular scale of this mismatch is small compared to the RGS LSF, so this hardly affects the modeled LSF.

The total RGS LSF width ($\approx 30$ arc seconds FWHM) combines the mirror PSF ($\approx 6$ arc seconds FWHM) and the
grating scattering and internal misalignment's. RGA scattering is the most dominant factor (see Fig. \ref{fig:lsf}).

\section{Detector calibration}
\label{sec:detcal}

Due to radiation damage, the detector properties change over time. 

The CCD response is only used to separate orders and has only an indirect
effect on the spectral resolution for a grating spectrometer. Its main impact is the data selections in CCD pulse height
that affect the instrument effective area.


\subsection{Detector model}

\begin{figure}

\resizebox{\hsize}{!}{\includegraphics [bb = 80 120 720 540] {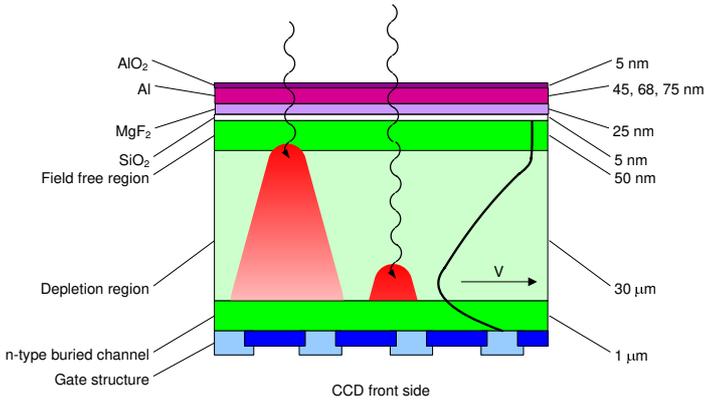}}
\caption{Model of the CCD indicating all different layers. X-rays entering from the top will be absorbed at a certain
depth. An X-ray photon of low energy might be absorbed close to the field free region, losing part of its
charge; higher energy X-rays will be absorbed closer to the gate structure and therefore the
charge diffusion in the depleted region will be smaller (indicated by the flat mountain shaped area
in the figure). At the right the potential in the CCD is indicated.}
\label{fig:ccd_cut}
\end{figure}

We developed a model for the detector response based on a simplified and phenomenological
approximation of the interaction of a photon and the detector.  
Fig. \ref{fig:ccd_cut} shows a crosscut through an RGS back illuminated CCD. The probability $P$ for absorption of
the X-ray photon at penetration depth $x$ is:

\begin{equation}
 \frac{\dd P}{\dd x} = \frac{1}{\tau} \cdot e ^ { -x / \tau }   \label{q:e1}
\end{equation}

\noindent with $\tau(E)$ the absorption depth for silicon for X-ray photon energy $E$ as derived from \cite{Henke}.

At the spot of the X-ray photon absorption, an amount of electric charge is released proportional
to the incident energy of the photon. We define the electric charge in units of the energy of the X-ray photon.
This scale of course implies the calibration of the electronic signal.

Not all charge released will make it to the output node of the CCD.
The collected charge $Q$ from the X-ray event depends on the depth at which the charge is generated. The deeper the
penetration, the closer to the front side structure and the more of the generated charge is collected. This is
modeled by:

\begin{equation}
 Q = \left\{ \begin{array}{ll}
	     0 & {\rm for~} E < T \\
	     T + ( E - T ) \cdot ( 1 - e^{-{(x/b)}^{1/3}} ) & {\rm for~} E \geq T \\
	     \end{array}\right.  \label{q:e2}
\end{equation}

\noindent with $T$ the threshold for charge detection and $b$ a scale parameter. These parameters were derived from
ground calibrations and set at $T=50$~eV and $b=20.95$~nm. 

The charge probability density can be written as:

\begin{equation}
 \frac{\dd P}{\dd Q} = \frac{\dd P}{\dd x} \cdot \frac {\dd x}{\dd Q} \label{q:e3}
\end{equation} 

From (\ref{q:e2}) we derive:

\begin{equation}
\dd Q = ( E - T ) \cdot e^{{-(x/b)}^{1/3}} \cdot {\left( \frac{x}{b} \right)}^{-2/3} \cdot \frac{1}{3b} \dd x
\label{q:e4}
\end{equation}

and:

\begin{equation}
\frac{\dd P}{\dd Q} = \frac{1}{\tau} \cdot e^{-x/\tau} \cdot \frac{3b}{E-T} \cdot
		     e^{{(x/b)}^{1/3}} \cdot {\left( \frac{x}{b} \right)}^{2/3} \label{q:e5}
\end{equation}

From \ref{q:e2} it follows:

\begin{equation}
 {\left( \frac{x}{b} \right)^{1/3}} = \lln \frac{E-T}{E-Q}  \label{q:e6}
\end{equation}

which allows to eliminate $x$ from (\ref{q:e5}):

\begin{equation}
\frac{\dd P}{\dd Q} = \frac{3b}{\tau} \cdot \frac{1}{E-Q} \cdot \lln^2 \left( \frac{E-T}{E-Q} \right) 
		      \cdot e^{-\frac{b}{\tau} \lln^3 \left( \frac{E-T}{E-Q} \right) }
\label{q:e7}
\end{equation}

To this function (\ref{q:e7}) a constant partial event floor $t$ is added that has constant probability density below
the incident energy and is zero above. The resulting redistribution function $\frac{\dd R}{\dd Q}$ thus
becomes:

\begin{equation}
   \frac{\dd R}{\dd Q} = \left\{ \begin{array}{ll}
                                 \frac{\dd P}{\dd Q}     & {\rm for}~ Q \geq E \\
                                 \frac{\dd P}{\dd Q} + t & {\rm for}~ Q < E
                                 \end{array}\right.	\label{q:e8}
\end{equation}

The partial event floor $t$ is normalized to the total partial event fraction $P_{\rm part}$:

\begin{equation}
   P_{\rm part} = \int_0^E t \: \dd Q = t \cdot E 
\end{equation}

\noindent $P_{\rm part}$ is 0.0348 for $E=1.9$~keV and scales with 
$(1 - e^{-1.832 \cdot E})\times \tau(1.9 {\rm keV})/\tau(E)$ with $E$ expressed in keV.

Finally, this shape is convolved with a Gaussian with a width of $\sigma$ to account for 
the fano \citep{Fano} and amplifier noise:

\begin{equation}
   \sigma = \sqrt{ ( a \cdot \sqrt{s} \cdot w )^2 + w f E }
\end{equation}

\noindent with $a=5.4$ the amplifier noise, $w=3.65$ the electron creation energy,
$f=0.14$ the Fano noise factor and $s$ a scale factor, depending on the OCB pixel size
with $s=1.4$ for 3x3 OCB and $s=2.5$ for 1x1 OCB.

\vskip 0.3 cm

\begin{figure}
\begin{center}
\resizebox{0.7 \hsize}{!}{\includegraphics [angle=90.0]{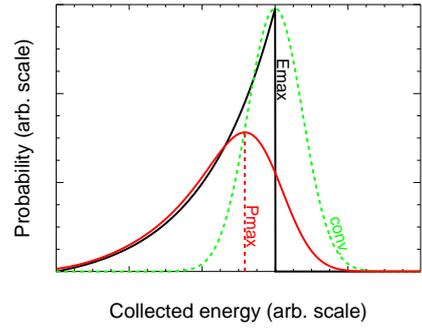}}
\end{center}
\caption{Schematic view of the effect of convolving the strongly asymmetric collected charge 
probability (solid black curve) for a photon with energy $E{\rm max}$ with a symmetric Gaussian shape (broken
green line) of the noise distribution. The resulting probability distribution (red curve) will have its maximum
$P{\rm max}$ shifted to lower energy than the original photon energy $E{\rm max}$.}
\label{fig:convtail}
\end{figure}

\begin{figure}
\resizebox{\hsize}{!}{\includegraphics [bb = 70 355 520 620] {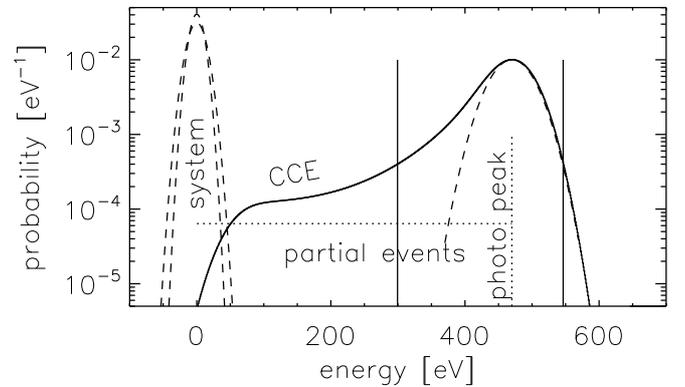}}
\caption{Model of the CCD response, or charge collection efficiency (CCE), which is the probability
for the collected energy, illustrating the key components.
Also shown is the 95 \%  window which is the default data selection (vertical lines). For completeness
we also show the typical read-out noise (system) which gives the typical output for a pixel without absorbed X-ray. 
The electronic bias level of the CCD has been subtracted from the energy axis, causing the read-out noise
system peak to extend below the zero energy level.}
\label{fig:ccd-model}
\end{figure}

\begin{figure}
\resizebox{\hsize}{!}{\includegraphics [angle=90.0] {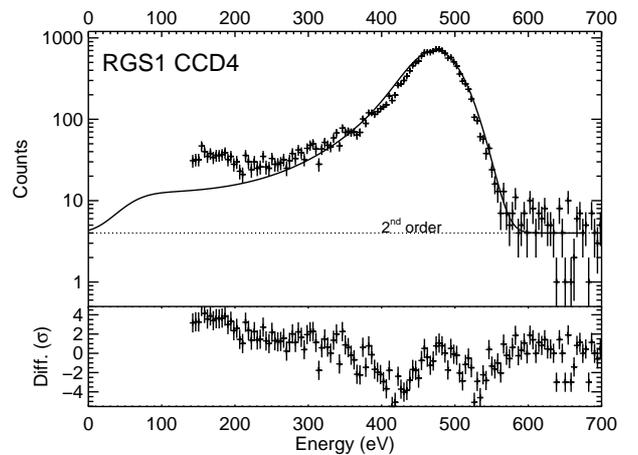}}
\caption{Modeled and observed CCD response for the \ion{N}{vii} \lya~ line (24.779~\AA = 500 eV). The data is a sum of 15
Capella measurements.}
\label{fig:overlay1}
\end{figure}

\begin{figure}
\resizebox{\hsize}{!}{\includegraphics [angle=90.0] {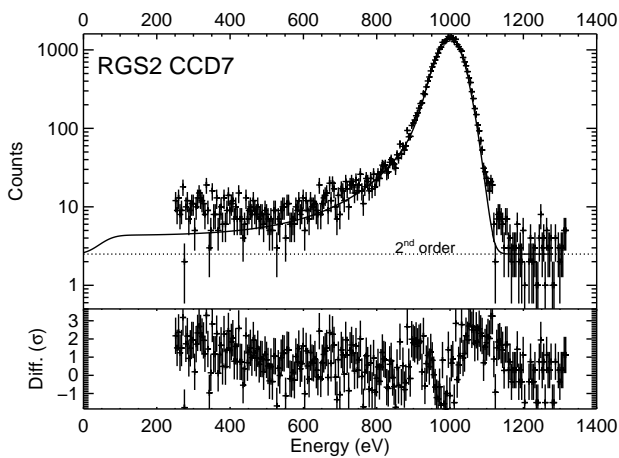}}
\caption{Modeled and observed CCD response for Ne X Ly-$\alpha$ line (12.132~\AA = 1022 eV). The data is the sum of 15
Capella measurements.}
\label{fig:overlay2}
\end{figure}

Fig. \ref{fig:ccd-model} shows the different components in the final model function.
This CCD redistribution function is checked in flight at regular intervals to see whether
the CCD charge collection is stable using bright mono-energetic emission lines of our wavelength calibration
sources (notably Capella) (Figs. \ref{fig:overlay1} and \ref{fig:overlay2}). The incomplete charge collection of charge
created near the back side (see Fig. \ref{fig:ccd_cut})  
which mainly affects low energy X-rays, or X-rays just above the Si edge, causes the tail of
the function to lower energies. 
This strongly asymmetric function with its tail towards low energies, combined with the convolution
by a symmetric Gaussian for
the Fano and amplifier noise, causes the peak of the function to be shifted towards lower energies,
as schematically shown in Fig~\ref{fig:convtail}, to below the
original incident X-ray energy. The effect is less at higher energies.
In both figures 7 and 8 the effect of
leakage of the second order events is shown. Also clearly visible is the improved resolution at higher
energies and the tail of the noise floor at lower energies.

This redistribution function can be regarded as the response to mono-energetic incident X-rays, in
which the peak of the collected energy does not necessarily coincide with the initially absorbed energy.
Using such a distribution enables calibrating a simple linear gain relation between collected charge
and CCD electronic output. When the observed peak of the redistribution function would be used to define
the CCD energy scale, a highly non-linear function would be needed.

\vskip 0.3 cm

The user can define any window for event selection. Generically the window is defined
in terms of percentage of the integrated CCD redistribution function at equal cutoff levels left and right of the  
maximum. The default selection uses 95\% of the events, identical for all energies. Fig. \ref{fig:ccd-model} shows this selection.
This nominal window is relatively insensitive to small variations in CCD gain and uncertainty in the
redistribution model. The CCD gain is calibrated at two yearly intervals by shifting
the observed CCD redistribution to the model distribution. The uncertainty in the gain 
calibration is estimated at $\leq 1$\%. The effect of the gain uncertainty on effective area is far
less than 1\%, since this is a second order effect.

Whereas the number of selected events, which are the actually detected photons, is largely insensitive to gain variations,
the effect of the visible/UV blocking filters as well as variations in the Si thickness are clearly
important. Si thickness variations
over the CCDs were measured by using the interference patterns of internal reflections of infrared light 
(1032 nm) within the CCD chip. Due to the production process the CCDs are thinner near the 
edges whereas the central part of each CCD is typically flat within 2 $\mu$m. The edges can
be up to 3 (one side) or up to 6 $\mu$m (other side) thinner which effects the QE at 1.8 keV of
at most -4\%. This effect is only relevant for those CCDs which cover the wavelength ranges between
1.5 and 1.82 keV (Si-edge) and is properly taken into account for the relevant CCDs in the calculated
efficiencies. 

In the model we have included absorption by the Al filter, deposited on top of the CCD 
with a varying thickness between 44 and 75 nm
and a MgF$_2$ isolation layer between 23 and 29 nm located in between the Al layer and the Si (see Fig. \ref{fig:ccd_cut}),
as measured by the CCD manufacturer (EEV). The uniformity of these layers was verified by UV/optical measurements
at 250, 406 and 540 nm.
Interpretation of these results was hampered by the lack of knowledge about the surface roughness
of these layers. The effect on total absorption of possible variations in surface roughness below 24 \AA{} is estimated
to be less than 1\% increasing to up to 9\% at the longest wavelength ($> 35$~\AA ). 


\vskip 0.3 cm

\begin{figure}
\resizebox{\hsize}{!}{\includegraphics [bb = 70 360 560 720] {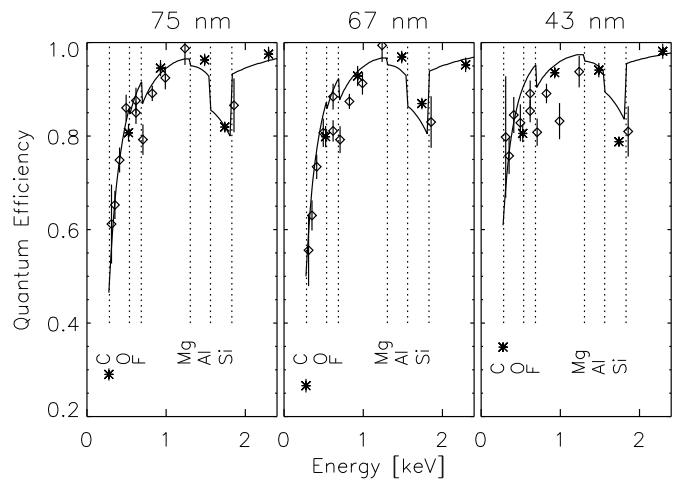}}
\caption{CCD quantum efficiencies, as measured during ground
calibrations at Panter \citep{herder}. The symbols indicate the measured data points. Stars: X-ray atomic line energies,
diamonds: monochromator data. The broken lines indicate the energies of atomic edges of the indicated atomic
species. The solid line is our model.}
\label{fig:ccd-qe}
\end{figure}

All these factors, which also include the signal selection thresholds, are taken into account to determine
the total QE of the CCD and these results were verified by ground measurements
(Fig.~\ref{fig:ccd-qe}).
The shown model is the calculated QE
based on physical (and measured) properties of the device as given in Table ~\ref{table:CCD}.
A fair match between the measured and calculated
QE is observed but, particularly at lower energies,
for the C, O and F sources
large deviations
could easily occur. This is mostly due to the fact that at low energy the natural
line width of these sources is relatively large. Attempts were made to use a separate monochromator to narrow the
natural line widths of the sources, but these results turned out to be less reliable due to 
unpredictable and uncalibrated inhomogeneities in the monochromator beam. 

\begin{table*}
\caption{
CCD parameters ranges
}
\label{table:CCD}
\begin{center}
\begin{tabular}{rll}
\hline
Al thickness CCD [nm]		& 44 - 75 	& depends on CCD position \\
MgF$_2$ thickness [nm]		& 23 - 29	& different per CCD  \\
Event size (3x3 OCB)		& 1.4  		& binned pixels \\
Individual CCD total central Si thickness [$\mu$m]	& 26.6 - 33.4 	& different per CCD \\
Si change of thickness over single CCDs [$\mu$m]	& 3 - 6 	& decrease of thickness towards CCD edges \\
\hline
\end{tabular}
\end{center}
\end{table*}

\subsection{CCD radiation hardness}
\label{sec:ccdrad}

Cosmic charged particles hitting the CCD detectors can damage the CCD. Either the Si depletion
region is damaged, giving rise to local low resistivity causing high dark current and hot pixels, or the charge
transfer layer at the front side (opposite the back side where the X-ray photons enter) is hit, which may cause
an increase in charge transfer inefficiency (CTI). Increased CTI will cause a drop in the PHA pulse height and a
decrease in the CCD energy resolution. When CTI becomes too high, $1^{\rm st}$ and $2^{\rm nd}$ orders cannot 
be separated any more and the dispersed spectrum will become indiscernible from the system peak.

A set of protection measures have been 
implemented. By applying proper shielding around the detector, 3 cm Al coated with a Au layer,
only particles with high energies, mostly protons, will reach the detector directly in addition to the secondary
radiation from these particles in the shielding. Also the design of the CCDs and 
their operations reduce the effect of radiation damage. Buried channels in the CCDs confine the charge
to a limited volume, reducing the relative size of the damage with respect to the signal. An
operational temperature range can be selected for which the dark current and cosmetic blemishes can
be reduced. Together these measures allow to handle  the
expected dose of $10^9$ 10 MeV equivalent protons and ~1 krad (after applying the 3 cm shield), which is representative 
for the expected dose over a mission of 10 year \citep{jansen1989}. 

However, Chandra experienced an unexpected
damage of their CCDs during the first part of the mission \citep{prigozhin2000,dell2000}. This could be
attributed to soft protons focused by the X-ray mirrors onto the detectors, encountered during the passage through
the radiation belts.  
To quantify this effect for our detector, flight representative RGS CCDs were subjected to protons 
at various low energies in the Van der Graaff accelerator in Utrecht. 
Table \ref{table:CCD_proton} presents the results for one representative CCD illuminated from
the CCD backside, which is the same side as illuminated in orbit.
The device was about 24 $\mu$m thick although
there is a significant variation over the CCD. We selected 6 different energies with a
stopping range from 10 $\mu$m to 30 $\mu$m. At low energies the protons are stopped in the 
bulk material whereas at the higher energies the protons are stopped in the layer where the charge is
transferred. At the highest energies the majority of the protons are stopped by the gate structure,  
or pass the CCD without causing serious damage.



\begin{table}
\caption{Sensitivity of RGS CCDs for soft protons normalized to a fluence of $5~10^{7}$ protons per cm$^{2}$
(stopping ranges from SRIM: http://www.SRIM.org).}
\label{table:CCD_proton}
\begin{center}
\begin{tabular}{lrrr}
\hline
$E_{\rm proton} $		& stopping range 	& CTI/transfer 	&  ${\rm{I_{dark}}}$  \\
MeV				& $\mu$m	& [$10^{-5}$] 	& ratio (post/pre)  \\
\hline
0.75				& 10.9		& 3.4 		& 6 \\
1.15				& 20.5   	& 7.8 		& 2 \\
1.27				& 23.8 		& 95.7 		& 5\\
1.40				& 27.7		& 306.6 	& 46 \\
1.45				& 29.2		& 61.6 		& n/a \\
1.55				& 32.4	  	& 18.0 		& n/a \\
\hline
\end{tabular}
\end{center}
\end{table}

\begin{figure}
\begin{centering}
\resizebox{\hsize}{!}{\includegraphics [bb = 80 370 540 680]{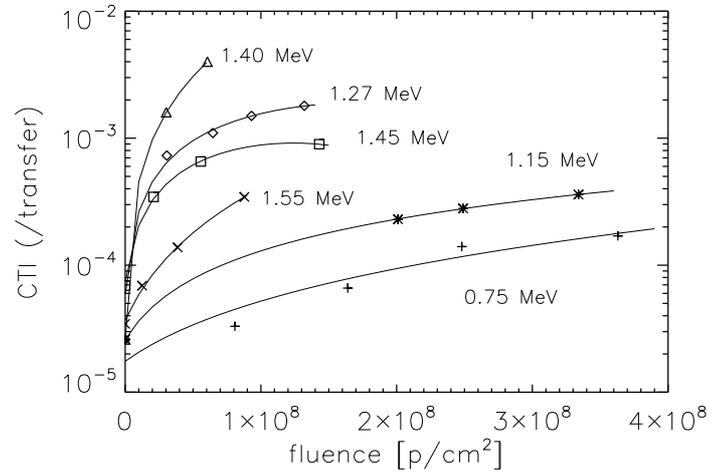}}
\caption{CTI as a function of the fluence for different energies of the protons. Curves are fit
to the data.}
\label{fig:CTI_soft}
\end{centering}
\end{figure}

\begin{figure}
\begin{centering}
\resizebox{\hsize}{!}{\includegraphics [clip]{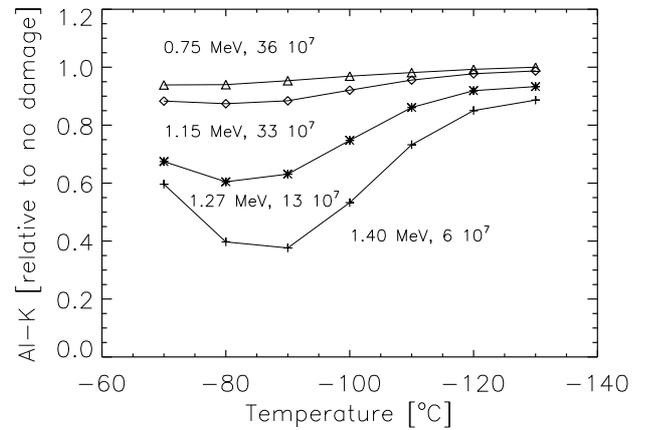}}
\caption{CTI as reflected in the PHA position of the Al-K line as function of temperature and radiation dose.}
\label{fig:charge_loss}
\end{centering}
\end{figure}

In Fig. \ref{fig:CTI_soft} the change in CTI is given for different radiation
levels and different proton energies. Although it was planned to test each energy up to a fluence of $10^8$ protons/cm$^2$ 
at the lower energies the beam was about a factor of two more focused than anticipated. This explains the differences in the
maximum fluences. As can be seen the CTI degrades most between 1.27 and 1.40 MeV, corresponding to a stopping range between 
24 and 32 $\mu$m, which is
in fair agreement with the thickness and uniformity of the CCD. In addition, we observed that the charge transfer
efficiency improves considerably if the CCD is stored at a higher temperature for one night following the irradiation
by the protons. This is a known
effect in CCDs. In Fig.~\ref{fig:charge_loss} 
we provide the integrated CTI (over typically 200 transfers) for 4 different energies at their
maximum fluence. As is also known from higher energy protons, the largest CTI occurs between -80 and
-100 $^\circ$C. At higher temperatures the CTI improves artificially; due to the high dark current the traps
are filled for a large fraction of the time. However, under this condition it is hard to impossible to 
separate the real events from the dark current. Cooling the CCDs below -100 $^\circ$C improves the
CTI considerably because the release time of charge from the traps becomes longer and thus their impact on
the CTI will be less (they are filled for an increasing part of the time). The curves in
Fig.~\ref{fig:charge_loss} are for a different total dose. No data were measured to correct for this at
other temperatures than -80 $^\circ$C. Finally we list in Table~\ref{table:CCD_proton} the increase
in the dark current. This also peaks at 1.40 MeV which is consistent with stopping of the protons
around the interface between the bulk Silicon and the gate structure. 

Based on these results, the energy dependence of the proton scattering from
the mirror and gratings (section \ref{sec:protscat}), and the spectral shape for soft protons ($E^{-1.7}$),
the RGS CCDs were expected to be hardly affected by soft protons. This has indeed been 
confirmed, although it is impossible to separate soft and hard protons using flight data only.
The absence of a filter wheel has not caused any unexpected problems. To identify a
time period of high proton flux the RGS detector closest to the optical axis (CCD9) is monitored and if the count rate for a 
part of the CCD not corresponding to the source image increases above a set level, the EPIC filter
wheel will be moved to its closed position. 


\subsection{ Read-out noise and dark current}
\label{sec:ccdnoise}
\vskip 0.15 truecm

\begin{figure}
\resizebox{\hsize}{!}{\includegraphics[angle=0, clip, bb =  15 15 300 120]{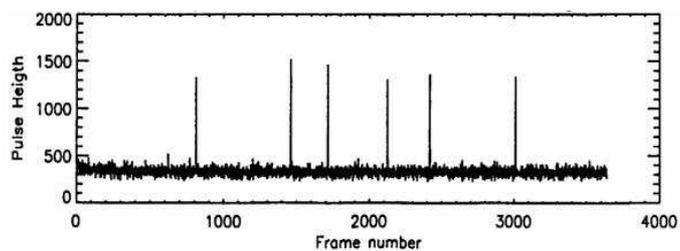}}
\caption{Response of a single detector pixel to X-rays}
\label{fig:real_pixel}
\end{figure}

In Fig.~\ref{fig:real_pixel} we show the typical response of a single detector bin when exposed to X-rays.
Most of the time no X-rays are observed, and the detected charge is a measure for the electronic noise and dark current. 
When an X-ray is absorbed in a bin the corresponding charge increases
(the spikes in the figure). The data free from X-ray events are called the noise floor or system peak 
in the RGS specific descriptions.

\begin{figure}
\resizebox{\hsize}{!}{\includegraphics [bb = 80 370 540 660] {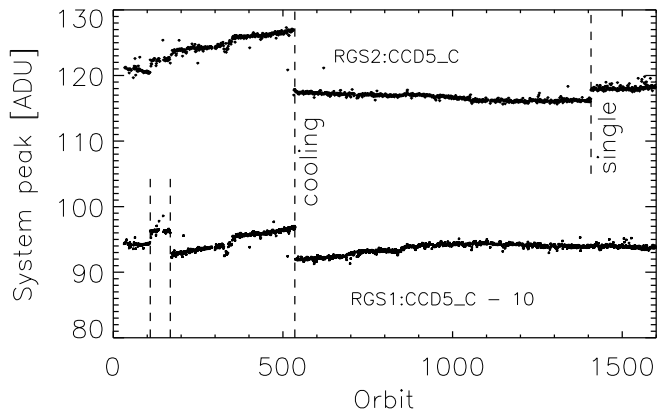}}
\caption{Evolution of system peak for one node of CCD5 in RGS1 and in RGS2 (other CCDs show a similar behavior). The
sudden increase of the peak around orbit 110 (most prominent in RGS1) is due to a solar flare, while the 
drop around orbit 168 (also indicated for RGS1) is due to the change in bias setting. Cooling lowers the system
peak around orbit 532, while the single node readout in orbit 1408 increases the RGS2 system peak.}
\label{fig:system_peak}
\end{figure}

The noise floor level is formed by the integrated dark current, on top of the electronic offset. 
Due to radiation damage and operational conditions, dark current and consequently the noise floor gradually
changes over the mission lifetime (Fig. ~\ref{fig:system_peak}). Major changes in the operational
conditions are indicated. Clearly the increase due to a strong solar flare (orbit 110), the change due to the CCD
 bias tuning (orbit 168), the reduction in dark current due to lowering the operational temperature (orbit 532 and 537), and
the increase in dark current following the change to a single node read-out which resulted in longer integration times, 
in RGS2 (orbit 1408), are visible. 
From the change in noise floor between the two node read-out and the single node read-out 
corresponding to a change of about a factor 2 in read-out time, the dark current could be
determined as 0.14 e$^-$/bin/s where a bin is 0.00656 mm$^2$.

\begin{figure}
\resizebox{\hsize}{2 in}{\includegraphics [bb = 90 370 570 670]{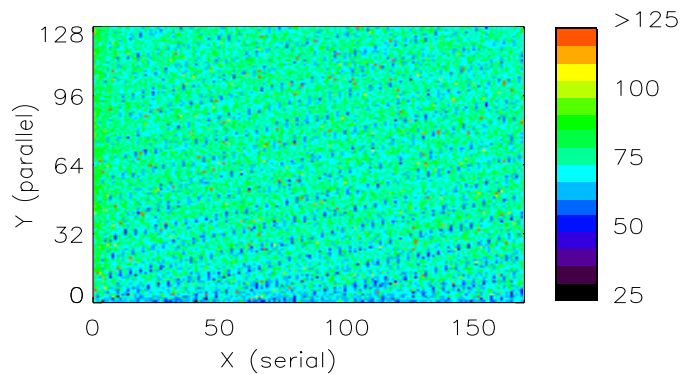}}
\caption{Fixed pattern noise (RGS2:CCD1\_C, two-node readout)}
\label{fig:fixed_pattern}
\end{figure}

In addition, we observe a small ($\Delta$ ADC $\leq$ 30 ADU) systematic pattern in the maps of the noise 
floors in the CCDs (Fig.~\ref{fig:fixed_pattern}). Following the change to the single node
read-out this pattern disappeared in RGS2, clearly indicating that this pattern is caused
by  cross talk between the two read-out chains. 
Although the effect is small compared to the energy resolution of the detectors ($\geq 100$ ADU), and affects
only $\approx 1$\% of all pixels it can, in principle, be corrected if the pattern is known. Unfortunately this
noise pattern is not fixed and drifts 
slowly over time and is not correlated with any of the instrument operational steps (e.g. 
there is no synchronization with the start of the read-out of a CCD). 

\begin{figure}
\includegraphics [width=3.0cm, angle=90.0, clip]{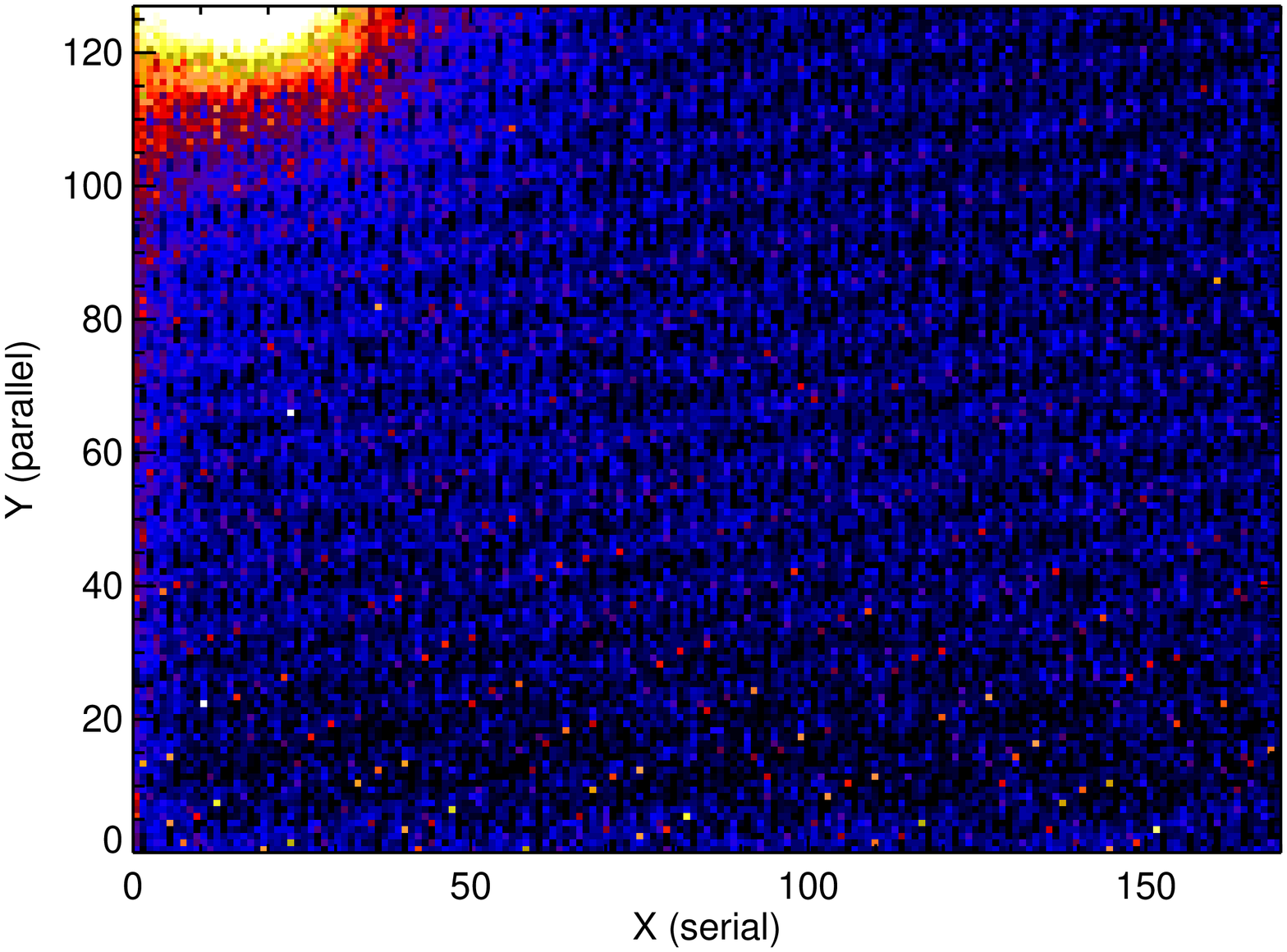}
\hskip 0.1 in
\includegraphics [width=3.0cm, angle=90.0, clip]{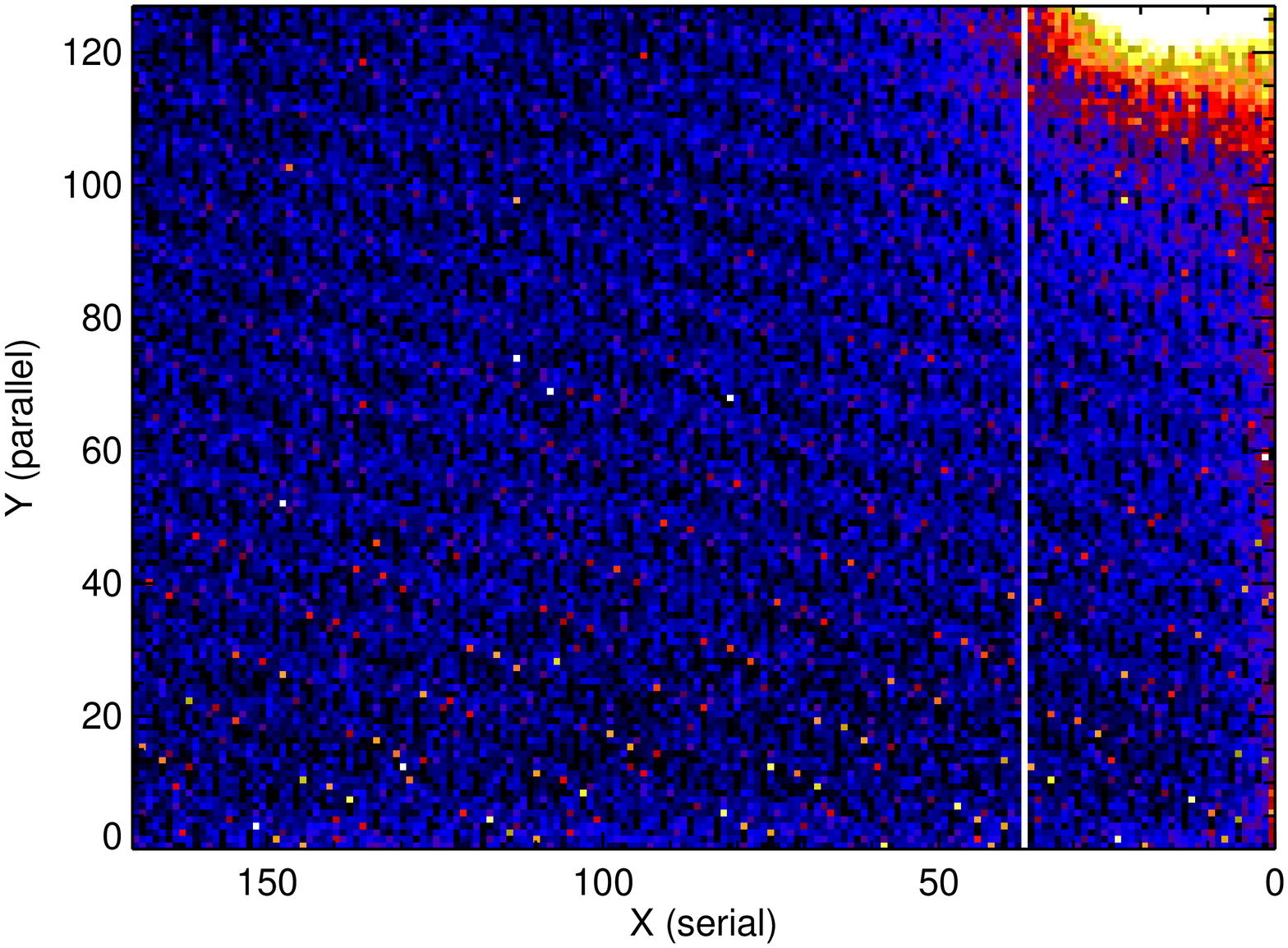}
\caption{Hot spots on RGS1 CCD1 with separate images of nodes C and D. The hot spot intensity
is about 180 ADC units above the system level floor of about 100 ADC units.}
\label{fig:hot_spot}
\end{figure}

Following orbit 786 RGS1 CCD1 developed two hot spots (Fig. ~\ref{fig:hot_spot}). The origin
of these spots is not clear as it shows up symmetrically in one CCD only. If it is radiation damage or a 
change in stress in the Si due to aging of the gluing to the detector bench, one
does not expect it to be symmetric for the two read-out nodes. 
It is possibly related to flat-band voltage shifts of the CCD due to 
radiation damage but this can not be verified. Fortunately this feature is limited to this single
CCD only and its area does not increase over time. The onboard software excludes
these areas from processing.

\vskip 0.3 cm


\subsection{Cosmetic defects}
\label{sec:defects}

\begin{figure}
\resizebox{\hsize}{!}{\includegraphics [angle=0, clip]{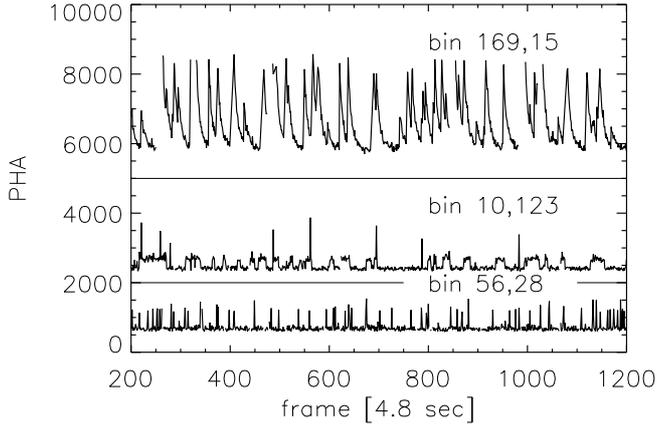}}
\caption{Variations in temporal behavior of cosmetic defects (orbit 0440, before the cooling). The
top two panels are offset by 2000 and 5000 for clarity. The small
gaps in the time sequence are due to the on-board processing where patterns which are not consistent
with the expectation for X-rays will be rejected to reduce the telemetry load.}
\label{fig:flickering_pixel}
\end{figure}

Local defects and damage by cosmic rays in the depleted silicon material can give rise to 
hot pixels and columns. 
Although the CCDs were selected based on the absence of hot pixels, the instrument
includes some CCDs with hot pixels and hot columns (see Table~\ref{table:hot}). 
The presence of these cosmetic blemishes is expected to vary over the life time of
the mission. Due to radiation damage new hot pixels or columns can appear, and also flickering 
pixels (also called random telegraph noise) can be expected. Due to the mobility of the originating 
charge traps some hot pixels may also disappear over the mission lifetime.
This is illustrated in Fig. \ref{fig:flickering_pixel} where the charges of a few noisy bins are shown.
Bins with a content below the lower threshold (e.g. only 
read-out noise and dark current) are rejected onboard and have a typical pulse height of 100 -- 120.
The variation in the patterns is huge, varying from fast damping to a somewhat bi-polar state. Following the 
cool down (see later) the effect of these defects in the crystal was significantly reduced.

\vskip 0.1 cm

Searching for these cosmetic blemishes has been fully automated and
we adopted an approach to reject all potential cosmetic blemishes. 
The signature of bad pixels and bad columns differs from the signature of astrophysical lines which cover more pixels. 
Therefore we compare the frequency that a given 
pixel or column is above the read-out noise with its neighbors, neglecting the actual pulse height
value of a given bin. If this frequency exceeds its expectations based on a binomial error propagation, 
the bin is rejected  
$ ( N > M + \sigma _p \sqrt{N(1-\frac{N}{F}) + M(1-\frac{M}{F})}$ with $N$ the total number of counts 
in a given pixel, $M$ the minimum of the total counts in the neighbouring pixels, $F$ the total number of frames 
and $\sigma_p$ is set at 5. For the rejection of hot columns we use a similar approach.

\begin{table}
\caption{Hot columns and pixels (3 x 3 OCB) defined in the calibration files; {\it suspect} pixels/segments
are pixels/segments which are rejected on the ground but are not persistent over
the mission lifetime. The orbit is the start of the validity interval of the corresponding 'BadPix'
calibration file (see table \ref{table:CCF})}
\label{table:hot}
\begin{center}
\begin{tabular}{llrl}
\hline
orbit	& CCD		& column 	& segment/comment \\
\hline
\hline
RGS1	&		&		&		\\
\hline
1	& CCD1\_D	&  38 		& hot column \\
	& CCD2\_C 	&  166 		& hot column \\
	& CCD3\_D	&  93 		& hot column \\
	& CCD5\_D	&  119 		& hot column \\
	& CCD7\_D	&  77		& hot segment 1-33 \\
	& CCD7\_D	&  74		& hot segment 1-33 \\
	& CCD9\_D	&  43 		& hot column \\
	& all		& 		& suspect: \\
	&		&		& 27 pixels, 5 segments, \\
	&		& 		& 5  columns \\
\hline
276	& CCD1\_D	&  38 		& hot column \\
	& CCD3\_D	&  93 		& hot column \\
	& CCD9\_D	& 43		& hot column \\
	& all		& 		& suspect: same  \\
\hline
536	& CCD1\_D	&  38 		& hot column \\
	& CCD3\_D	&  93 		& hot column \\
	& CCD9\_D	& 43		& hot column \\
	&	all	& 		& suspect: \\
	&		&		& 5 pixels after cooling  \\

\hline
543	& CCD1\_D	&  38 		& hot column \\
	&	all	&		& suspect: 5 pixels \\
\hline
1110	& CCD1\_D	&  38 		& hot column \\
	&		&		& suspect: 4 columns \\
	& CCD1\_C 	& 1 - 31	& suspect segment 121-128	\\
	& CCD1\_D 	& 1 - 31	& suspect segment 121-128	\\
\hline
1417	& CCD1\_D	&  38 		& hot column \\
	& CCD1\_C 	& 1 - 31	& hot segment 121-128	\\
	& CCD1\_D 	& 1 - 31	& hot segment 121-128	\\
	& all		&		& suspect: 4 columns \\
\hline
\hline
{\bf RGS2 } &		&		&	\\
\hline
1	& CCD9\_C	& 94		& hot column \\
	& CCD9\_C	& 147		& hot segment \\
	& all		&		& suspect: 1 column \\
	&		&		& and 20 pixels \\
\hline
532	& CCD9\_C	& 94		& hot column \\
	& all		&		& suspect: \\
	&		&		& 1 pixel and 1 segment\\
\hline
1110	& CCD9\_C	& 94		& hot column \\
	& all		&		& suspect: 7 columns \\
\hline
\end{tabular}
\end{center}
\end{table}

\begin{figure}
\resizebox{\hsize}{!}{\includegraphics[angle=90.0]{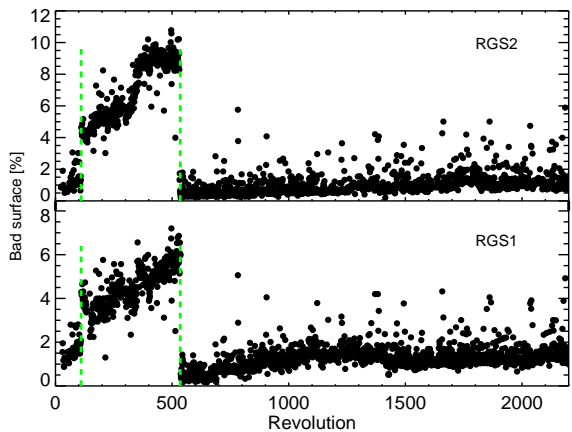}}
\caption{Bad area per detector as detected by the standard monitoring tools. The first increase corresponds to a
major solar flare in orbit 110, the
drop around orbit 535 is due to the cooling of the detectors from -80 $^\circ$C to -110 $^\circ$C. Individual outliers
are due to peculiar observational circumstances, like e.g. periods of very high background, observations of very bright
sources, or very short observations.}
\label{fig:rgs_bad}
\end{figure}

The  approach to reject cosmetic blemishes is rather conservative, but as long as it is properly taken into account in
the 
computed exposure for each observation,
it ensures that hot pixels/columns do not cause any artificial emission lines. 
This method translates into
an exposure map with large pixel to pixel variations affecting the effective
area curves (Fig.~\ref{fig:aeff}). 
In Fig.~\ref{fig:rgs_bad} we display the 'dead' area of the two detectors 
over the lifetime of the mission. The step-wise increase in the early part of the mission corresponds
to solar activity. Following the cool down of the detector the number of hot items reduced by a factor
5 -- 10 and we expect that the total affected area will stay below 4 \% over the next 10 -- 15 years, depending of course on the
actual solar activity in the future.

\subsection{Warm/cool pixels}
As remarked before (section \ref{sec:defects}), defects in the depleted region may cause local higher dark current,
or extra charge traps, leading to hot and cool pixels respectively.
Hot pixels are discarded, but many pixels will only show a slightly elevated dark current, which is added to the
charge of recorded X-ray events, potentially shifting them outside of the pulse height selection area.  
Cool pixels will lower the recorded charge of events.
To mitigate this effect, averages are made of CCD full frame images, recorded during three consecutive orbits around
the orbit of the observation being processed. These averages are used to subtract the system peak level on a pixel by
pixel bases, which will correct for the different system level peaks of hot and cool pixels.


\subsection{Charge Transfer Inefficiency}

Ionizing particles from cosmic rays can also cause traps in the Si lattice structure at the layer where
the charge is shifted from pixel to pixel to the output node.
These traps are characterized by the trapped charge relative to 
the total charge packet, the probability that the charge
remains trapped during the charge transfer, and the probability that 
the charge will be emitted during the time between charge packets. 
Although this effect depends on the time
since the last charge transfer took place through the pixel, in first approximation
we can describe this by an average Charge Transfer Inefficiency (CTI) that neglects
the pixel to pixel variations and pixel history. The unit of CTI is defined as the fractional loss
of charge per physical charge transfer on the CCD. On the RGS CCDs there are up to 1024 transfers in the CCD
X-direction, in single node readout, for the most distant pixel to reach the readout node. 
Our general CTI model is justified because the
CCD pulse height distribution is only used to separate the orders and not for the spectral 
response of the instrument; our CTI correction does not need very high accuracy. The
uncertainty in the total CTI correction for a pixel corresponds to 1\%
in the total charge for an average CTI per transfer of $10^{-5}$, to keep it the same
order of magnitude as the gain error. 
A wider selection of the CCD response function results in a smaller error, but leads to more background.
In addition to traps, the CTI is also affected by the voltages applied to the clocks when
shifting charge through the CCD. 
Since we know the energy of the incident
photons from the position on the detector, we can tune the voltages for the clocks and/or calculate 
the CTI corrections.

\vskip 0.1 cm

After failure of the RGS1 CCD7 readout chain CCD clock voltages were reduced to limit stresses
on the components (orbit 168, table \ref{tab:events}). These clock voltage reductions were further tuned  
for RGS2 CCD2 which was done in orbits 192 and 537.
and the clock bias of RGS1 CCD2 needed adjustment in orbit 1389.

\vskip 0.1 cm
 
The Charge Transfer Inefficiency is calculated independently for the parallel transfers (transfer of
the image section to the storage section or of a row in the storage section to the read-out row) and
for the serial transfers (transfer through the read-out row toward the read-out amplifier).

For the parallel transfers the CTI is deduced from the measurement of a strong point source (Mrk421)
at different positions in the cross 
dispersion direction. The difference in collected charge for different positions in the
cross dispersion direction is only due to the CTI loss, as the deposited energy is
given by the grating dispersion equation. This can be simply converted into a CTI calibration model. 

For the serial transfers a slightly different approach is followed. 
CTI during readout causes the collected charge to drop depending on the number of charge
shifts, given by the CCD column. Using the known energy of the photons from the dispersion
relation of the grating, the relative change in charge deposited by the incoming photons over the CCD is known.
The additional charge loss per transfer gives the CTI model.

The parallel and serial CTI are small (few times $10^{-5} $). Pre-launch values were around
$ 0.9\times10^{-5}$.  Every two years special calibration 
measurements are executed to calibrate the CTI. Additional calibrations were performed immediately
after cooling of the RGSs and following the change from double to single node readout of RGS2. 
Table~\ref{table:cti} shows a 
moderate increase of the CTI of $10^{-5}$ over 1000 orbits but the CCD to
CCD variations are larger (a few $ \times 10^{-5} $. These parameters can be
determined to about $ 0.5 \times10^{-5} $ which
corresponds to an uncertainty of the energy loss of $ \lessapprox 0.5 \% $.

\begin{table*}
\centering
\caption{Average CTI degradation per camera as retrieved from the calibration files (CCF) and variation (maximum
difference) over all CCDs, in between brackets}
\label{table:cti}
\begin{tabular} {l r r r l}
\hline
RGS	&	orbit	& serial CTI $ (10^{-5})$ & parallel CTI $ (10^{-5})$ & comment \\
\hline
RGS1	&	30	& 3.45 (2.90) 	& 0.94 (0.05)	& pre launch data\\
	&	260	& 4.16 (2.61)	& 4.26 (1.60)	& first measurement in orbit \\
	&	536	& 5.05 (4.30)	& 4.07 (2.00)	& post cooling \\
	&	807	& 5.07 (6.53)	& 6.55 (4.10)	& \\
        &      1397	& 4.89 (3.83)   & 7.68 (4.41)   & \\
	&      1839	& 4.96 (3.53)	& 8.92 (4.98)	& \\
	&      2021	& 5.92 (5.42)	& 11.13 (3.70)	& \\
\hline
RGS2	&	25	& 3.29 (3.14) 	& 0.95 (0.05)	&  pre launch data \\
	&	260	& 3.55 (5.51) 	& 3.05 (2.98)	&  first measurement in orbit\\
	&       532	& 3.87 (2.96)	& 2.14 (2.81)	&  cooling to $-115^\circ{\rm C}$ \\
	&       536	& 3.48 (2.82) 	& 2.42 (2.38)	&  cooling nominal \\
	&	807	& 3.17 (2.65)	& 6.11 (3.73) 	&	\\
	&      1405	& 2.86 (1.17)	& 8.06 (3.71)	&  to single node readout  \\
	&      1839	& 2.71 (1.04)	& 9.67 (4.96)	& \\
	&      2021	& 2.58 (1.82)	& 11.09 (5.29)  & \\
\end{tabular}
\end{table*}

\begin{figure} 
\resizebox{\hsize}{!}{\includegraphics[angle=0, clip, bb = 90 360 540 620]{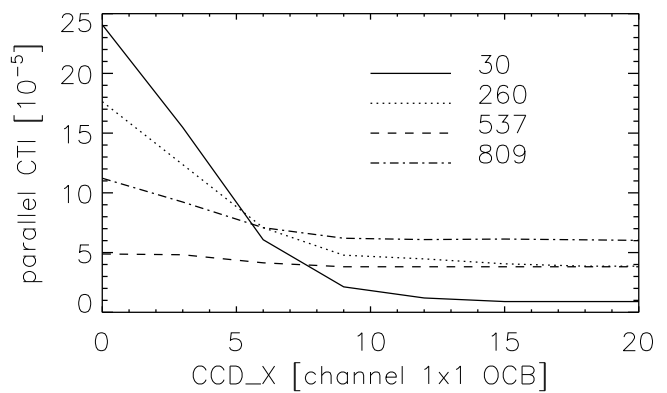}}
\caption{Average parallel CTI for RGS1 near the edges for different orbits. The uncertainty towards
the edge increases towards about
$5\times10^{-5}$. For channels $>10$ the uncertainty is of the order of $1\times10^{-5}$}
\label{fig:cti_parallel}
\end{figure}

Near the edges of the CCDs the parallel CTI is significantly larger. 
To reduce the dead space between adjacent CCDs in the detector array, the CCDs were 
sawn at the two sides, causing additional stress and hence defects in the Si. (Fig.~\ref{fig:cti_parallel}).
During the early part of the 
mission the parallel CTI near the CCD edges is $ \sim 20 \times 10^{-5} $, and it
reduces to a few times $ 10^{-5} $ after the cooling of the camera in orbit 539. Over time the
CTI increases again.

\begin{figure}
\resizebox{\hsize}{!}{\includegraphics[angle=0, clip, bb = 81 375 530 665]{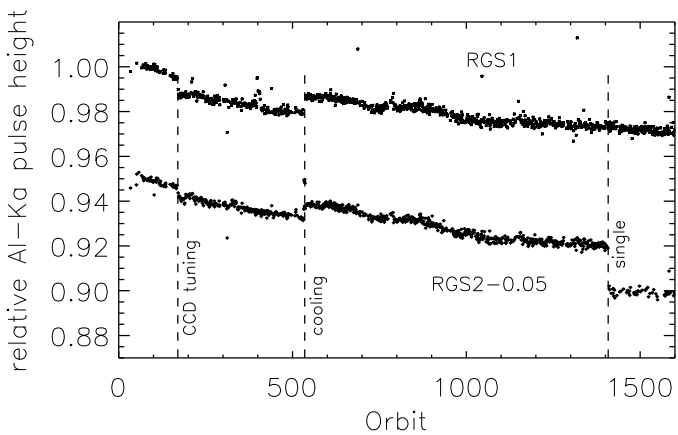}}
\caption{Change in charge transfer efficiency as reflected by the gradual decrease of the uncorrected pulse height value
for events from the onboard Al-K$\alpha$ calibration sources. 
Major events: tuning of the CCD settings, cooling of the detector and switch to the single node read-out for RGS2, cause
sudden shifts in CTI. Occasional outliers are caused by peculiar observations like e.g. bright sources, observations with high background and
very short observations.}
\label{fig:cti}
\end{figure}

The change in CTI was independently verified by comparing the position of the onboard calibration line
positions as function of orbit. This is shown for the Al-K$\alpha$ line in Fig.~\ref{fig:cti} where the
position of the Al-K onboard calibration source is shown before any gain and CTI correction. 
Apart from the jumps associated to CCD tuning, cooling and change to single node readout,
a gradual decrease in pulse height of 3 \% over the lifetime of the mission is observed.
This corresponds to a change in CTI of $10\times10^{-5}$ over 1600 orbits, consistent with the change of
$5\times10^{-5}$ in 800 orbits for the higher CCD channels, as shown in Fig. \ref{fig:cti_parallel}.

\vskip 0.1 cm


\subsection{Optical sensitivity}

\begin{figure}
\resizebox{\hsize}{!}{\includegraphics[angle=90, clip]{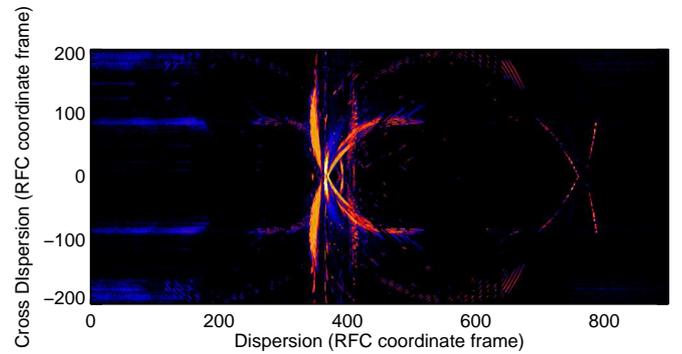}}
\caption{Simulated image for a source at -1.5 degree off-axis in the direction of increasing
incident angle on the grating. The projected photons near Z=400 (the narrow lines) are double reflections
from the mirror. The fan like shape is due to single or double reflections from the mirror and a single reflections 
from the grating. 
In the RFC coordinate frame position (0,0) is de center of CCD5 which is the position of the grating 
first order blaze wavelength. The coordinates are in units of 0.01 mm.
 }
\label{fig:optical_sens}
\end{figure}

The mirrors and gratings are near perfect reflectors for UV/optical light and this
results in additional charge on the detectors for optically bright stars. 
The energy assigned to the X-ray photons increases by a corresponding amount and therefore is incorrectly
interpreted. On the EPIC cameras this is known as "optical loading". The
filter wheel in front of the EPIC instruments allows the user to reduce the optical 
light. For RGS, with its elongated focal plane, a different solution is used. For sources within the field of view
the specular reflected photons are imaged on the ${\rm zero}^{\rm th}$ order position of the grating,
40 mm away from the CCD bench, and they do not interfere with the dispersed X-ray spectrum. Because optical radiation
has much longer wavelengths, the higher
orders of optical light will end up far beyond the CCD bench. However, light from bright
off-axis sources can interfere with the recorded spectrum (Fig.~\ref{fig:optical_sens}).

\begin{figure} 
\resizebox{\hsize}{!}{\includegraphics[angle=0, clip, bb = 60 370 560 700]{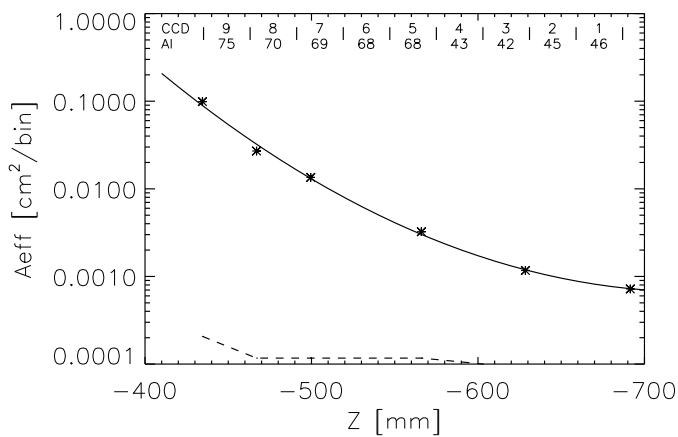}}
\caption{
Global stray light sensitivity of RGS as modified by the varying Al
filter thicknesses (in nm). Separate CCDs and their Al filter thicknesses are indicated (top). 
The sensitivity for focused (solid line) and non focused light (dashed line) is given separately. The asterisks
indicate ray-trace model calculations for the indicated positions on the CCDs.
}
\label{fig:straylight}
\end{figure}

Based on simulations we calculate the corresponding effective area for different off-axis
angles and position on the detector. This is shown in Fig. \ref{fig:straylight}. 
Based on this sensitivity we verified whether possible contributions from the sky would affect the 
RGS response. As expected contributions from bright stars dominate the sensitivity, relative to 
the night glow features and diffuse night sky. Very bright stars near the field of view can
be avoided by proper scheduling, as the optical sensitivity depends only on the incident
angle with respect to the gratings. Without further scheduling constraints to m$_v$ $\leq$ 5, 4\% of the observations
would be optically contaminated.
Also the population of weaker field stars adds to the optical contamination.
Reduction of the optical load is required to keep the contribution of this specular reflected star light to less
than 1 electron per readout. This reduction
varies from 10$^{-4}$ near the optical axis to 10$^{-2}$ for the CCDs at the largest distance to the optical
axis and has been achieved by an Al layer deposited directly on the CCDs. To compensate for
the difference in sensitivity as function of the dispersion direction, we used a  
thicknesses of 75 nm for CCD9 and CCD8, 68 nm for CCD7, CCD6 and CCD5, and 45 nm for the other CCDs.

\begin{figure} 
\resizebox{\hsize}{!}{\includegraphics[angle=0, clip, bb = 90 370 570 670]{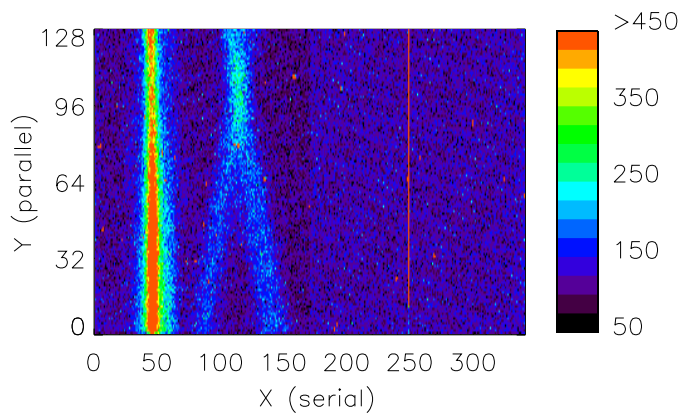}}
\caption{Recorded intensity distribution on CCD 3 with Canopus at -2 degree off axis}
\label{fig:canopus}
\end{figure}

The suppression of UV/optical light was verified in orbit by observing Canopus an F0 star with m$_v$ = --0.72
between --3 and +2 degree off axis along the dispersion direction. The observed image (Fig.
\ref{fig:canopus}) has the correct shape and its intensity is within a factor 2 from
the model. Considering the 
approximations used (a 1 nm change in Al filter thickness may change the transmission up to  15\%)
this is reasonable and the observation strategy of an avoidance angle for bright
sources is sufficient to neglect the effect of optical/UV light on the detector response. 

\vskip 0.3 truecm


\subsection {Stability detector response}

\begin{figure}
\resizebox{\hsize}{!}{\includegraphics[angle=0, clip, bb= 80 370 545 700]{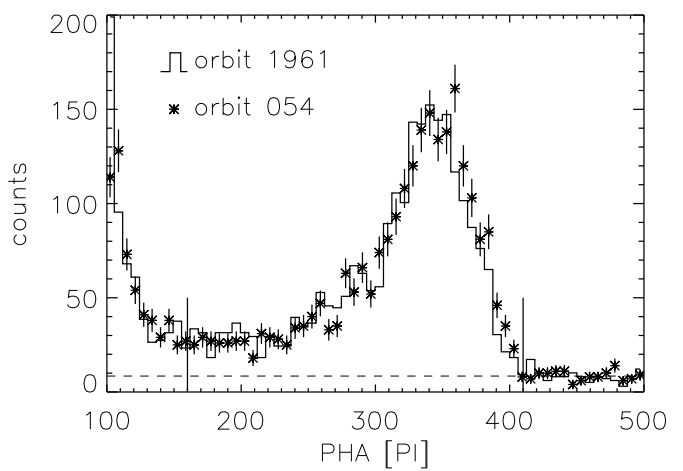}}
\caption{In-orbit CCD response at $\beta$ = 0.0715 radians (C VI line) for Capella observations in orbit 54 and 1961. The errors are only shown for orbit 54.}
\label{fig:ccd_redis}
\end{figure}

After applying all correction factors for the evolution of the detector response (CTI, gain, hot columns, 
read-out noise and dark current and the evolution of hot pixels and columns) we expect that
the calibrated CCD response is accurate over the lifetime of the mission.
For a detector behind a 
dispersive element this can be easily verified as the incident energy of each photon is known. 
In Fig. \ref{fig:ccd_redis} we show the calibrated CCD redistribution function
for one energy at the start of the mission and after 10 years in operations.

The stability of the detector response, after applying the
appropriate corrections is very good and the effect on the effective area is
negligible ($< 0.25 $ \%) assuming that the proper pulse height selection windows $\geq$ 95 \%
are applied.


\section{Gratings}

\subsection {Grating  model}

The grating response describes the intensity distribution of monochromatic radiation as function
of outgoing angle $\beta$ and its component perpendicular to it. 
The two major contributions are the flatness and the scattering properties
of the gratings. Also the grating to grating alignment is
a major component. 

From scalar diffraction theory one can calculate the total integrated scatter (TIS) out of the line core 
\citep{kahn1996}. This is given approximately by:
\begin{equation}
       TIS ~=~ k^2~\sigma^2 ~( {\rm sin}\, \alpha  + {\rm sin}\, \beta )^2 
\label{q:tis}
\end{equation}
where $ k = 2 \pi / \lambda $ is the wave number of the radiation and $ \sigma^2 $ is the mean square deviation
of the surface which follows from a simple analytical function that is characterized by $\sigma$ and a
scale length $ l $. This expression (\ref{q:tis}) is valid in the small amplitude limit where 
$ \lambda >> \sigma \, {\rm sin}(\alpha) $ and where the deflection angles of the surface are small
compared to the critical graze angle
for reflection. We get a good
description of the grating response using two components: an incoherent large scale
component with a correlation length of $1~\mu$m and an rms of 8 to 14 \AA{} depending on the grating master
and a coherent short scale component with a correlation
length of 0.29 $\mu$m and an rms of 15~\AA. This second component is due to a sinusoidal wave on top
of the groove structure and has been verified by Scanning Tunneling Micrograph images. 
%

For the diffraction efficiency we use an exact approach which solves the vector electromagnetic equations
in the space above the grating subject to the periodic continuity relations imposed by the groove surface. 
Based on synchrotron measurements we verified the blaze angle, which is the angle where the efficiency
peaks.

%

\subsection{Gratings calibrations}

Calibrations of the gratings included two steps: (a) verification of the grating response model
at a synchrotron and (b) measurements of the efficiency of each grating separately at a few
wavelengths.

\vskip 0.3 truecm

\begin{figure}
\resizebox{\hsize}{!}{\includegraphics[clip, bb= 70 370 540 705]{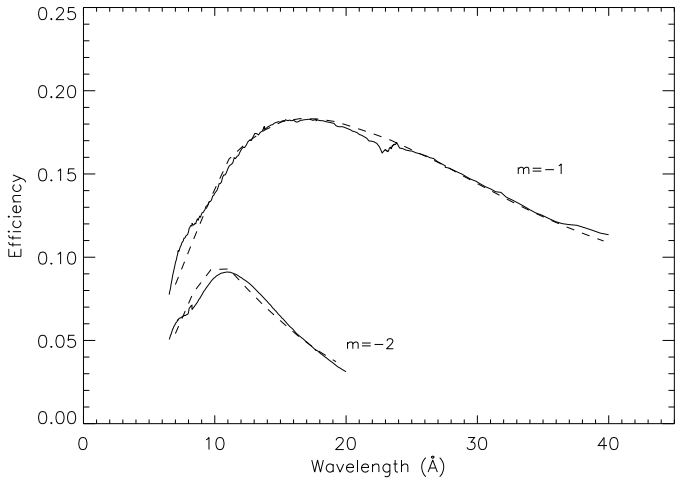}}
\caption{Efficiencies for first and second order as measured (solid curve) and as modeled (dashed
curve), for a single grating \citep{herder}.}
\label{fig:RGA_response}
\end{figure}

{\it The single grating response} was verified by measuring the  
efficiency of a single grating  at the Bessy synchrotron facility in Berlin 
(Fig.~\ref{fig:RGA_response}). Clearly there
is good agreement, allowing the blaze angle and the scattering contributions as free parameters. 
Near the oxygen and carbon edges a thin layer of hydrocarbons on the surface of the grating is not
included in our model, causing small differences. At short wavelengths the model, based on a scalar approximation of the 
diffraction theory \citep{Cottam1998} breaks down. 

\vskip 0.3 cm

{\it All gratings} were calibrated as part of
the screening of replicated gratings: their efficiency was measured at 8.34 (Al-K$\alpha$), 
13.34 (Cu-L$\alpha$), 17.59 (Fe-L$\alpha$) and 23.62 \AA~(O-K$\alpha$) and scattering measurements were
performed at the Al-K line. For a few gratings additional measurements of the inter-order scattering
were carried out. The  selection of flight gratings was based on a grading scheme where poorer candidates 
were rejected on the basis of their $m=-1$ efficiencies and the replica flatness (for RGS1: $\epsilon_{\rm Al} > 0.09 $ or 
$\epsilon_{\rm Cu} > 0.15 $ and flatness $< 15 \: \mu$rad and for RGS2: $\epsilon_{\rm Al} > 0.085 $ or 
$\epsilon_{\rm Cu} > 0.15 $ and flatness $< 9 \: \mu$rad). 


\vskip 0.3 cm

{\it The integrated reflection grating assembly} was tested at the Panter facility at MPE. 
This test included two components: checking the grating assembly by partial illumination of the
RGA and measuring the LSF (mirror + gratings) at some energies. 

The partial illumination was achieved by  blocking the 
mirror entrance aperture except for a small section, using the so-called Gl\"{u}cksrad. Following these
tests we concluded that apparently the reference point for this RGA was not set properly during assembly. 
This meant that for the first RGA we had to modify the orientation of the RGA with respect to the mirrors
by a change in  angle ($\Delta \alpha$ = + 4.56 arcmin and $\Delta x$ = 7 mm). In addition a rotation in the
cross dispersion direction, around the Z-axis, was
applied to reduce the scattering from the supporting ribs on the gratings ($\Delta \theta$ = + 1.71 arcmin). 
For the second unit such modification was not required. 
Following these modifications the alignment of the gratings was tested at the Panter long beam facility at
MPE and these were consistent with the expected results.

The full verification of the LSF is described in section \ref{sec:lsf} where we show the results for the
calibrations of the mirror and grating combination.

\subsection {In-orbit grating response verification}

The grating response in the dispersion direction is not easily verified in flight as
it includes a convolution with the incident spectral shape of the observed
object. When the spectral continuum has a strong gradient, or strong lines are present, spectral intensity will spill
to neighboring energy bands and mix with the grating response when the spectrum is integrated in cross
dispersion direction along the dispersion axis. 
A major factor in the 
grating response is the scattering by the grating due to irregularities and surface 
roughness of the grating (see Fig. \ref{fig:lsf}).

\begin{figure}
\resizebox{\hsize}{!}{\includegraphics[angle=0, clip, bb= 40 320 550 590]{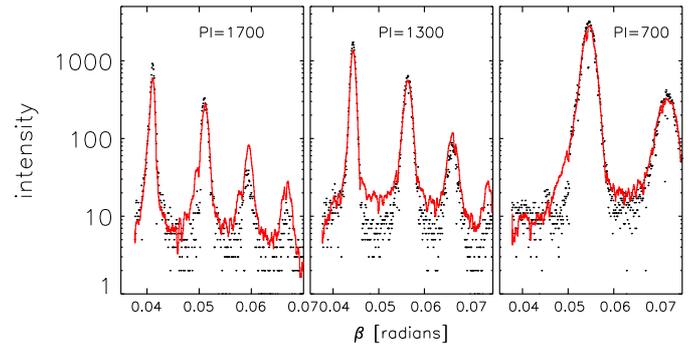}}
\caption{
Modeled and measured  $\beta$ distributions for given energies 
expressed in CCD pulse height (PI), where 1 ADU $\approx$ 1 eV. This corresponds to a horizontal slice through the
PHA-Beta plot in Fig~\ref{fig:data_space}. The peaks correspond to the different orders, starting from order -1 on the left
of each plot.
}
\label{fig:scatter1}
\end{figure}

\begin{figure}
\resizebox{\hsize}{!}{\includegraphics[angle=0, clip, bb= 40 320 550 590]{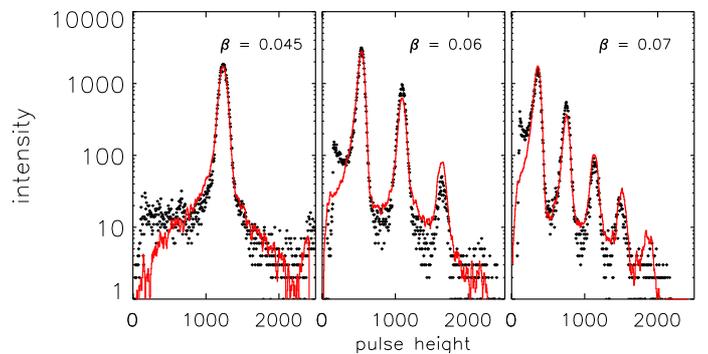}}
\caption{
Modeled and measured CCD pulse height distributions expressed in CCD pulse height (PI), where 1 ADU $\approx$ 1 eV,
for given $\beta$ angles in radians. This corresponds to vertical slices in the PHA-Beta plot in Fig~\ref{fig:data_space}.
The peaks correspond to the different orders, starting from order -1 on the left
of each plot. 
}
\label{fig:scatter0}
\end{figure}

The grating response was first verified using Mrk 421 for a number of
selections in the dispersion angle $\beta$ and in pulse height (Figs. \ref{fig:scatter0} and \ref{fig:scatter1}).  
The spectrum as a function of dispersion
angle ($\beta$) clearly shows the different orders and the scattering from the gratings in between
the orders. 
There is good agreement between the data and the modeled response for the first and second
order after applying the scattering component. 

\begin{figure*}
\resizebox{\hsize}{!}{
\includegraphics[angle=90, clip, bb=85 102 530 709]{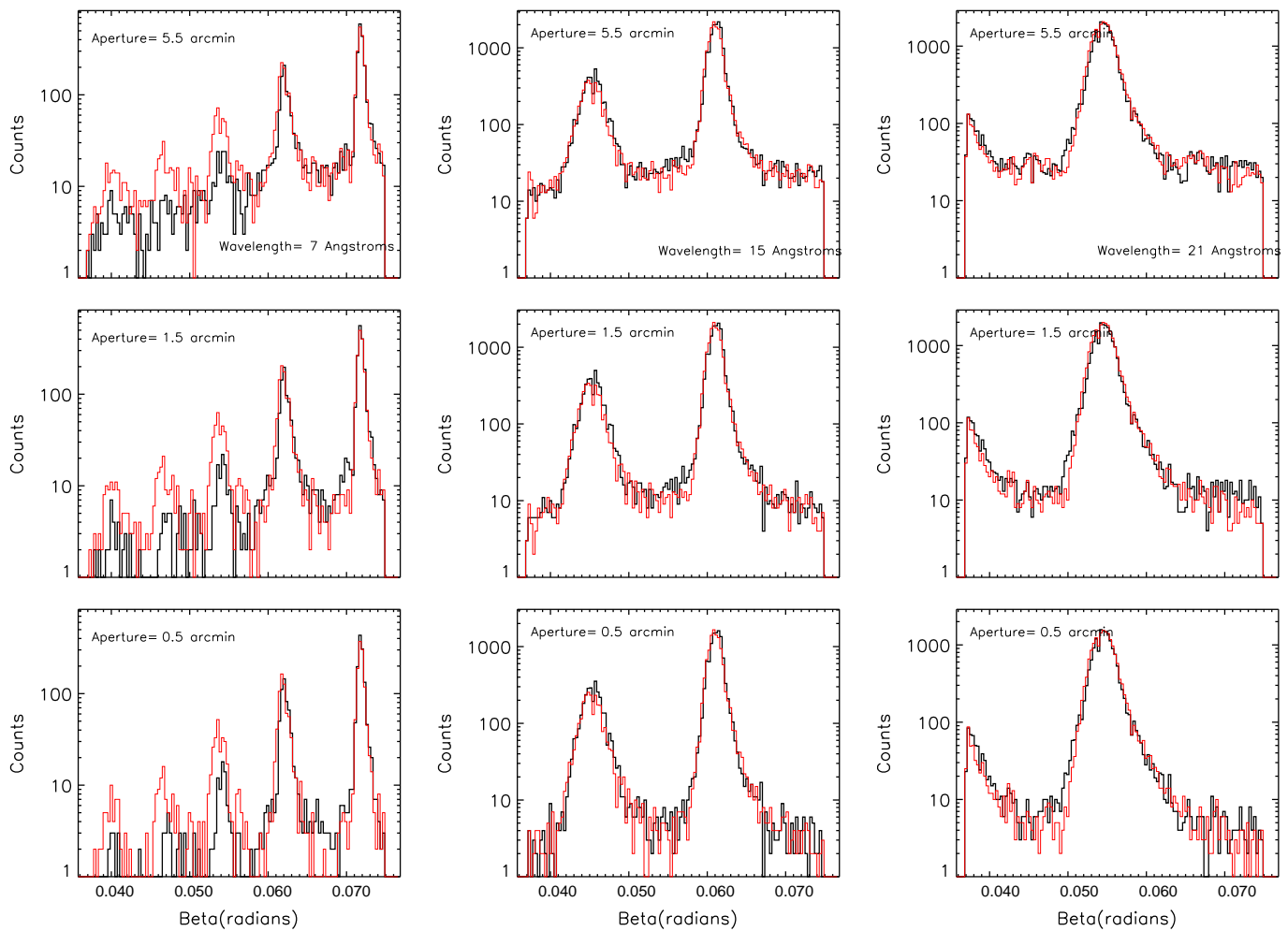}
}
\caption{Comparison of modeled response (red line) with observations (black line) for different cross
dispersion apertures (ranging from 0.5 to 5.5 arcmin, bottom to top), shown as intensities versus dispersion angle
(Beta) for different CCD pulse height selections (columns left to right) at 7, 15 and 21 \AA, or 1.7, 0.8 and 0.6 keV.}
\label{fig:scatter2}
\end{figure*}

This scattering also occurs in the cross-dispersion direction. 
In the two-dimensional data slice of
cross-dispersion versus the dispersion, or spectral axes,
the cross dispersion profile is affected by the scattering of neighboring spectral bin intensities.
However, its fractional amplitude is suppressed by a factor of $\sim 20$
compared to the scattering amplitude along the dispersion direction due to the difference
in relative incidence angle between the dispersion and cross-dispersion direction. When the spectrum is collapsed on the
cross dispersion axis the cross-dispersion profile is largely a convolution of the mirror PSF with the
grating response. 
Fig. \ref{fig:scatter2} shows the verification of this scattering component. Overall, good agreement
is observed, confirming the validity of our scattering model.

\subsection{Scattering of low energy protons}
\label{sec:protscat}
The low energy protons encountered by \xmm{} during its orbit form a potential risk for the performance of 
the CCD detectors (see section \ref{sec:ccdrad}). These protons are scattered and focused by the mirrors
onto the \xmm{} instruments \citep{aschenbach2007}. For the RGS an extra reflection by the gratings
is required in order for the protons to hit the CCDs. To estimate the efficiency of this process, 
a sample grating was put in a soft proton beam at the Harvard university, Cambridge accelerator for materials
science \citep{rasmussen1999}. It was found that for protons of 1.3 MeV, the CCDs would acquire per ${\rm cm}^2$
only $10^{-5}$ part of the radiation dose on the mirror entrance 
due to scattering off the grating.

It was concluded that for the RGS the required extra reflection by the gratings, in combination with
the lower vulnerability of our back illuminated devices for these soft protons, offers sufficient
protection. In principle the instrument can even be left in operational
mode during the periods of high radiation caused by soft protons.

\section{Line Spread Function}
\label{sec:lsf}

\begin{figure}
\resizebox{\hsize}{!}{\includegraphics{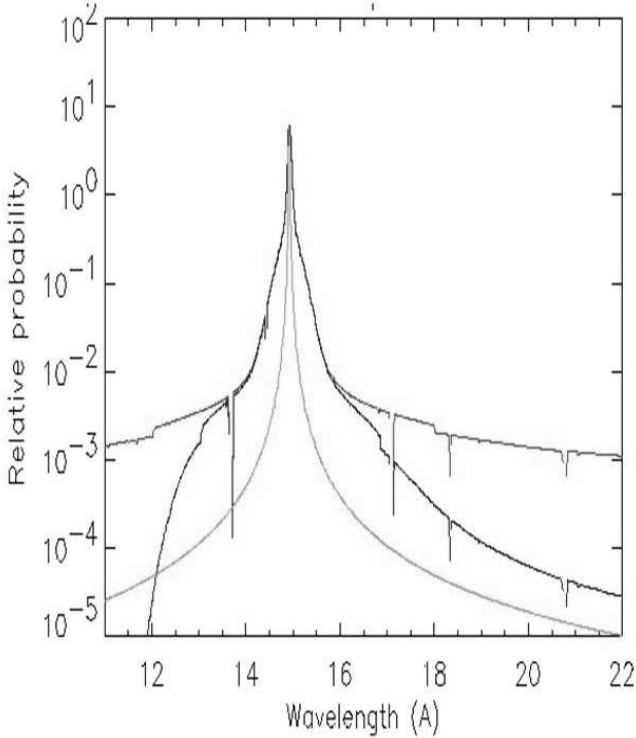}}
\caption{Modeled line spread function showing the different contributions of the mirror (narrow peak), convolved with the
small angle scattering (middle line) and wide angle scattering (top line).}
\label{fig:lsf}
\end{figure}

The line spread function (LSF) is determined by folding of the mirror response with the grating response, taking into
account the effect of data selections in the detector. This is schematically illustrated in Fig.~\ref{fig:lsf} where
the different components are shown. The sharp symmetric Lorentzian profile describes the mirror response.
After folding with the grating response the distribution shows a narrow angle scattering component (Gaussian
shoulders in the distribution) and additional large angle scattering wings. The downward spikes in this spectrum
are due to discarded hot columns. The small step function is due to the limited range over 
which certain scattering components are computed.
Beyond this range this component becomes insignificant, and only the total wide angle scattering flux is needed to compute the
total amount of flux lost from the center of the distribution, to get the proper effective area. 
Applying the data selection of the CCD the shape of this distribution becomes asymmetric in the wings.

The LSF was verified
in various stellar sources
(Capella, Procyon, HR1099, AT~Mic and $\lambda$~And). The instrument behaves slightly better than predicted 
by the folded responses, in the sense that the line peak is slightly narrower. This is modeled by a small ($\approx 1$ m\AA{}
equivalent for $1^{\rm st}$ order) manual modification of the figure parameters of the grating description
(see LINESPREADFUNC ccf calibration file). 

\begin{figure}
\begin{center}
\resizebox{\hsize}{!}{\includegraphics [angle=90, clip] {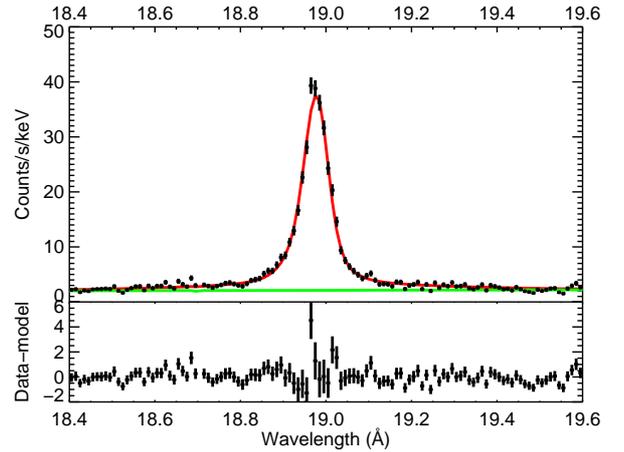}}
\caption{\small Measured (points) and modeled line spread function (line) for the \ion{O}{viii} \lya~ line
in HR 1099 (18.97 \AA), for RGS2. The assumed stellar continuum background is indicated by the lower (green) line.
The modeled line is shifted by 5~m\AA, to fit the measured line position. 
This 5~m\AA\ shift is within the wavelength
uncertainty and of the same order as the radial velocity shifts within the system which are between +40 and -65~km/s
\citep{Bopp1976}, or about $\pm3$~m\AA.
The difference between the model and the data is shown in the bottom plot.}
\label{fig:science_lsf}
\end{center}
\end{figure}

A typical example of the LSF response is shown in 
Fig.~\ref{fig:science_lsf}. The estimated contribution due to the background and the
stellar continuum causes the largest uncertainty
in the modeled wings of the line.

\begin{figure}
\vspace {-0.3 cm}
\begin{center}
\resizebox{\hsize}{!}{\includegraphics[angle=90]{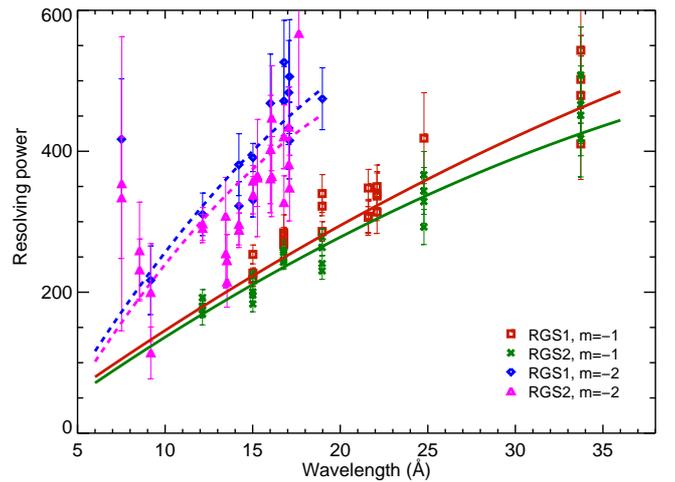}}
\caption{Resolving power for RGS1 and RGS2 implemented in the calibration given by the response matrix (lines), and measured from 
individual line profiles (points), showing the spread around the model, and demonstrating that the line widths follow
the trends defined by the physical model of the instrument.}
\label{fig:resolution}
\end{center}
\end{figure}

Using  relatively clean emission lines (\ion{Ne}{x} \lya, \ion{O}{viii} \lya,
 \ion{Fe}{xvii}, \ion{N}{vii} \lya, and \ion{C}{vi} \lya) we determined the resolving power ${\rm E}/\Delta{\rm E}$ for 
the two orders in the different instruments (Fig.~\ref{fig:resolution}). The resolving power
is between 100 and 600 in the first order and is two times
better in the second order.

The modeled
LSF and the observed LSF agree well after application of a correction factor for the internal alignment
of the gratings of RGS1. 
Based on the spread of the measured resolution for a given wavelength (Fig. \ref{fig:resolution}) astrophysical
line broadening of more than
$\Delta\lambda = 6 $ m\AA{} or 10 \% of the line width of the instrumental profile can
be detected significantly for strong lines 
(${\rm S/N}\gtrsim 25$).

\section{Wavelength calibration}

\label{sec:wavcal}
On the ground a full calibration of the wavelength scale with X-rays was not possible as the
mirror--grating--detector combination has to be at different positions due to the finite
source distance at the Panter facility at MPE. Therefore the wavelength calibration had to be performed in 
orbit.

\subsection{Wavelength scale}

Using a set of well defined emission lines the relative alignment of the three components,
mirror, gratings and detector, has been determined. We have used relatively clean
 lines (Table~\ref{table:lines}) in a number of coronal sources during the performance verification
phase: Capella (orbit 042, 046, 054), HR 1099 (orbit 025, 028, 031, 036) and AB Dor (orbit 091). In 
this analysis the barycenter corrections and
corrections for the velocity of the spacecraft are included. Since SAS version 13 these 
corrections can be applied (but not by default) by the standard software analysis system.

\begin{table}
\caption{Lines used to calibrate the wavelength scale and LSF. The last two columns indicate which lines
have also been used for the separate RGS1 and RGS2 LSF calibrations.}
\label{table:lines}
\begin{center}
\begin{tabular} {l l l l}
\hline
line			&	$\lambda_{\rm Laboratory}$	& RGS1 LSF & RGS2 LSF  \\
\hline
\ion{Mg}{xii} \lya	& 8.41		& no	& no 	\\
\ion{Ne}{x} \lya	& 12.134  	& no	& yes 	\\
\ion{Fe}{xvii}		& 15.015	& yes	& yes	\\
\ion{Fe}{xvii}		& 16.777	& yes   & no 	\\
\ion{O}{viii} \lya 	& 18.969	& yes	& yes	\\
\ion{O}{vii} {\it w}	& 21.602	& yes	& no	\\
\ion{O}{vii}		& 22.101	& yes	& no  	\\
\ion{N}{vii}		& 24.781 	& no	& yes 	\\
\ion{C}{vi} \lya	& 33.736	& yes	& yes 	\\
\hline
\end{tabular}
\end{center}
\end{table}

The dominant factors in the wavelength scale are (a) actual pointing position relative to the source, (b) a rotation of the 
RGA around the Y-axis (changing the incident angle) and (c) a linear shift in the Z-direction (dispersion). 
This is an over constrained problem because the final shift in wavelength depends on a linear combination
of  ${\rm sin}\,\alpha $ and ${\rm sin}\,\beta$. We have chosen a solution where we fixed the incident angle on
the gratings to the preflight calibrations, and modified the offset pointing (boresight) and the shift of the detector
(810 and 490 $\mu$m for RGS1 and RGS2 respectively).

\begin{figure}
\begin{center}
\resizebox{\hsize}{!}{\includegraphics[angle=90, clip, bb=61 68 540 717]{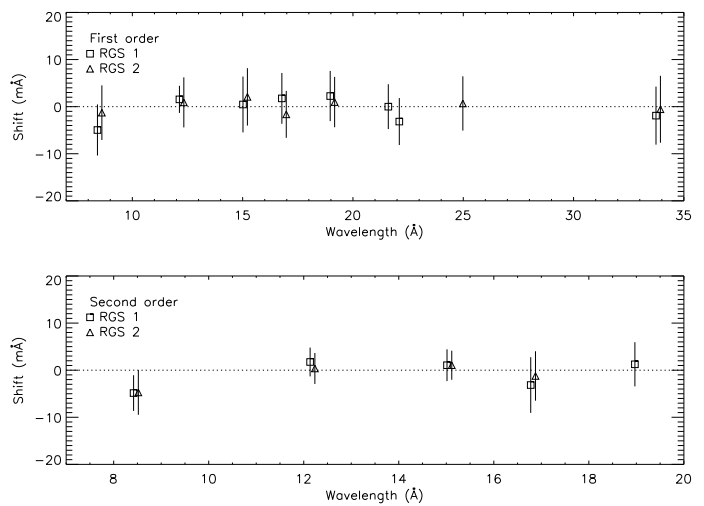}}
\caption{Difference between expected and observed wavelengths as a function of 
expected wavelengths for RGS1 (squares) and RGS2 (triangles). 
RGS2 data have been slightly (horizontally) displaced for clarity.
Error bars are the standard deviations of the statistics of the line fits. The spread of the data points
seems somewhat smaller than the error bars indicate. There is a possibility this is due to correlations between
spectral bins because of rebinning of the original CCD pixels, which might cause the statistical error
to be somewhat overestimated.
}
\label{fig:fig_wavelength}
\end{center}
\end{figure}

It was noted that the variation between observed and theoretical
line positions was larger than expected based on the built accuracies of the instrument. 
Bending of the telescope 
tube, the structure between the mirror platform and the detector, could have caused this variation. However,
this was ruled out because the focus of the mirrors, as recorded by the EPIC instrument, is perfectly stable. 
After collecting a large set of observations this variation turned out to depend 
on the angular distance between the Sun and the spacecraft pointing direction, the "Sun angle" (SA).
Considering that the mirrors are 
mounted at the intersection plane between the paraboloid and hyperboloid, a difference in temperature of 
less than 0.5 degree between observations with a different SA can explain a shift of up to 2~m\AA{}. 
This is well within the required temperature stability of the satellite  of 1 degree 
for the mirror and grating combination. This effect is included in the calibration since version 13 
of the SAS. In Fig.~\ref{fig:fig_wavelength} the accuracy of the wavelength scale is illustrated. 
Clearly there is no wavelength dependency after the calibration has been applied. 

Recently it was discovered that there appears to be a seasonal pattern in the \xmm{} pointing accuracy, with
an amplitude of 1--1.5 arcsec. This 
bore-sight variability, originally noted in the EPIC instruments, is thought to be caused by star-tracker instabilities.
These offsets are taken into account since September 2012.

\subsection{Wavelength accuracy}

\begin{figure} 
\begin{center}
\resizebox{\hsize}{!}{\includegraphics[angle=90, clip, bb=62 69 538 704]{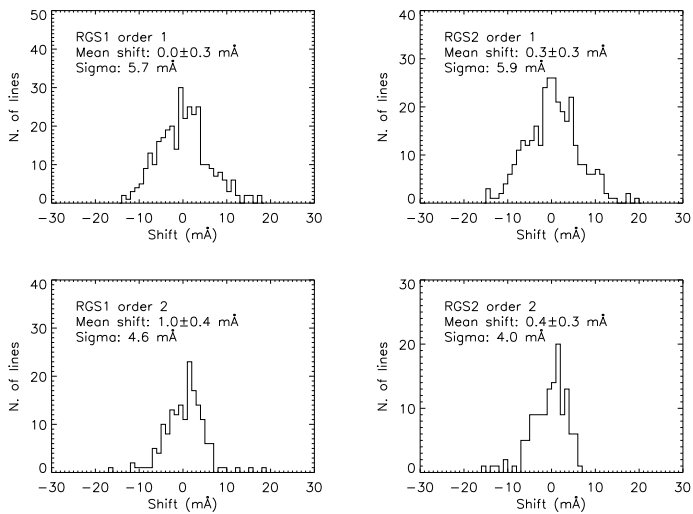}}
\caption{Distribution of the difference between the measured and the theoretical 
wavelength for nine emission lines in 56 observations of four stars (Capella, AB Dor, 
HR 1099 and Procyon).}
\label{fig:fig_wave_histo}
\end{center}
\end{figure}

To verify the ultimate accuracy of the wavelength calibration we have used the same lines for a
large number of observations 
of 4 coronal sources and calculated the distribution of the difference between the 
expected and measured positions for the first and second order lines (Fig. \ref{fig:fig_wave_histo}).
Table \ref{tab:wavacc} shows the accuracies and line shifts obtained for the lines used in the calibrations.
Applying all known corrections, shown in the last column of Table \ref{tab:wavacc},
line shifts are zero with a spread for first order of 5~m\AA.
Thus for any observed line
we expect to know its position with an accuracy of 5 m\AA{} (1 $\sigma$) corresponding to typically 75 km/s, at 20~\AA.

Further improvements can be made when lines are observed in both first and second order. The angle of incidence
on the grating for the particular observation can be independently solved
by forcing the corresponding first and second order lines to have the same observed wavelength.

The pointing accuracy of the \xmm\ telescope is 1.5 arc seconds, which corresponds to about 3 mA in the dispersion
coordinate. In addition, the emission lines of most coronal sources, due to the violent processes in the coronae
and/or the existence of companion stars have velocity uncertainties of 40--50 km/s or about 3 mA. The 
accuracy of our current wavelength calibration is very close to this theoretical limit of about 4.5 mA.

\begin{table}
\caption{Wavelength accuracy in m\AA{} for the applied corrections}
\label{tab:wavacc}
\centering
\begin{tabular}{cccc}
spectrum & a & b & c \\ \hline
RGS1 order 1 & $4\pm8$ & $2\pm7$ & $0\pm6$ \\
RGS2 order 1 & $9\pm8$ & $7\pm7$ & $0\pm6$ \\
RGS1 order 2 & $2\pm5$ & $1\pm5$ & $1\pm5$ \\
RGS2 order 2 & $4\pm5$ & $3\pm4$ & $0\pm4$ \\
\end{tabular}
\begin{minipage}{7cm}
\begin{enumerate}[a)]
\setlength{\itemsep}{0pt}
\item Fixed Bore-sight, no further corrections
\item Fixed Bore-sight + barycenter and stellar velocity corrections
\item Variable Bore-sight + barycenter and stellar velocity corrections and Sun angle (SA) corrections
\end{enumerate}
\end{minipage}
\end{table}

\vskip 0.1 truecm

\section{Effective area}
\label{sec:effarea}

\subsection{Effective area model and ground calibrations}

\begin{figure}
\resizebox{\hsize}{!}{\includegraphics[bb = 90 360 550 700]{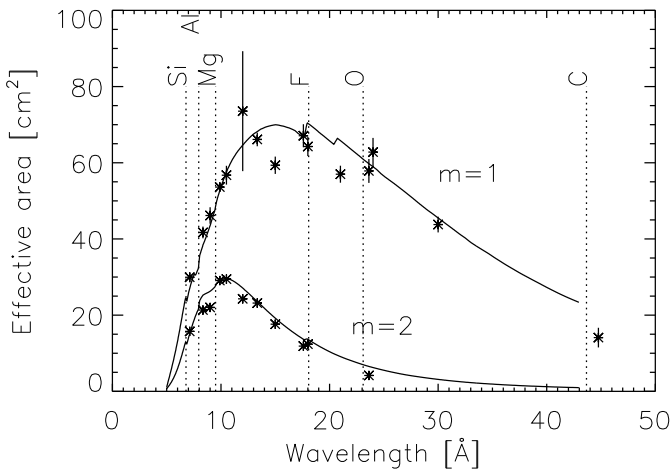}}
\caption{Effective area verified pre flight by end-to-end measurements at the Panter facility \citep{herder}.
This can be compared
with the measured in flight effective area as verified on celestial sources (Fig~\ref{fig:aeff}). The model is indicated by
the solid line. Sharp features are caused by either atomic absorption edges or differences between different CCD
Al layers. The broken lines indicate the relevant atomic edges. The asterisk symbols show the measured data points.}
\label{fig:eff_area}
\end{figure}

The effective area is a combination of the mirror effective area, the
grating efficiency and scattering and the detector quantum efficiency together with
any data selection applied. In Fig.~\ref{fig:eff_area} the results of the end-to-end tests
at the Panter facility are displayed including the expected 
effective area. The data are 
corrected for the finite source distance.

\vskip 0.3 truecm
These results indicate a 10\% accurate calibration of the effective area
over the 10--25~\AA\ range. At shorter wavelengths there is a steep gradient
in the grating efficiency, resulting in a significantly larger uncertainty of the effective
area. At the longest wavelength the CCD QE is less accurately known, up to 40 \%
(see Figure \ref{fig:ccd-qe}).
The longest wavelength point in Fig.~\ref{fig:eff_area} is well
beyond the maximum wavelength of 38~\AA\ for the in flight configuration of the RGS instruments. 

\skip 0.3 truecm

\subsection{In orbit scaling}

\begin{figure}
\resizebox{\hsize}{!}{\includegraphics[angle=90, clip]{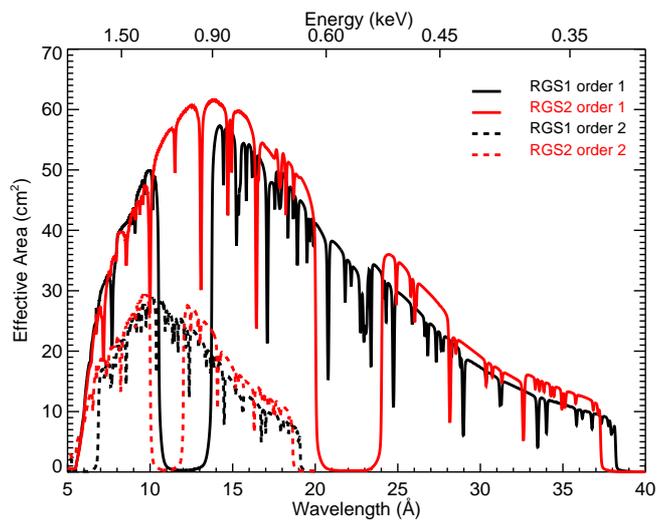}}
\caption{Effective area for RGS1 (black) and RGS2 (red) (two orders) as calculated by the SAS, based on in flight 
calibrations on celestial sources. Apart of the two failing
chains (CCD4 in RGS2 and CCD7 in RGS1) the CCD boundaries can be easily identified. In addition the effect of hot
pixels and columns can be seen.}
\label{fig:aeff}
\end{figure}

Considerable effort was invested to improve the effective area in orbit by 
observing continuum  sources with good statistics. A total of 1 Ms
observations of Mrk~421, PKS~2155-304 and other
selected sources (e.g. Vela pulsar, Sco X-1 off axis) are used. 
The resulting area result is shown in Fig~\ref{fig:aeff}.
All modifications of the area model are needed due to either uncertainty in the
model response or items not fully included in the calibrations. 

\vskip 0.1 cm

An unexpected  {\it vignetting} of the beam was deduced by comparison of the RGS1 and RGS2 response. After 
launch it turned out that the effective area for RGS1 was less than that for RGS2 and that it scaled
with the angle $ \beta $ of the dispersed rays ($A_{\rm eff}(RGS1)/A_{\rm eff}(RGS2) = 1$ for
$\beta=0.036$ and 0.8 for $\beta=0.072$). 
This dependency is consistent with a partial blocking of the 
through beam between the mirrors and the RFC and is explained by a 180$^\circ$ rotated baffle of the
grating assembly around the optical axis. The purpose of this baffle is to isolate the RGA thermally from its environment.
The exit opening is asymmetric with respect to the mirror optical axis to allow an unblocked view to the RGS camera. 
Although this should have been checked during integration, from a manufacturing point of view rotation
over 180$^\circ$ around the optical axis cannot be excluded. The RGS1 effective area is corrected for this using a linear
function of $\beta$.

\vskip 0.1 cm

The {\it QE} of CCD2 in RGS2 was modified based on continuum spectra of Mrk~421 and PKS~2155-304.
In the initial data an unexpected jump in the spectrum was observed for CCD2 in RGS2. 
The observed loss of QE can be explained and is modeled with an additional 40 nm layer of ${\rm SiO}_2$ on the CCD. 
This layer does match the observed absorption in both first and second orders. Why such an extra layer of ${\rm SiO}_2$
should exist on this CCD however, is not clear, and there may be another (unknown) cause for the observed loss in QE.

\vskip 0.1 cm

{\it Systematic flux variations} (up to 10\% of the flux) on scales of a few {\AA}ngstroms, as function
of wavelength were observed for sources
which are expected to show smooth, power law spectra (e.g. PKS~2155-304, Mrk~421). These deviations
from the expected power laws are attributed to unaccounted variations of the effective area.
To correct these flux variations, we took high statistics observations of Mrk~421 in orbits 84, 259, 546, 720 and 1084
and fitted absorbed power law spectra to the observations in the interval from 10 to 25~\AA. For each spectrum different
power law indices and normalizations were allowed, but only one ${\rm N}_{\rm H}$ column density for the interstellar absorption.
Thus the source was allowed to vary between observations, but the interstellar absorption remained constant. The same
set of parameters was fitted to both RGS1 and RGS2 spectra. It is assumed that
in the interval between 10 and 25~\AA{} the effective area of the RGSs is known best, and that in this RGS wavelength
interval Mrk421 shows a smooth power law spectrum.

Next the fitted power law and absorption model parameters were frozen and $12^{\rm th}$ order Chebychev
polynomials (which could follow the few \AA{} scale variations) were fitted over the full 5 to 38~\AA{} 
interval to the fit residuals for each RGS separately, 
but identical for all observations. These Chebychev polynomials are now used as corrections on our effective 
area and are included in the SAS calibration files.

\vskip 0.1 cm

\begin{figure}
\begin{center}
\resizebox{\hsize}{!}{\includegraphics [angle=0, clip, bb = 80 360 550 690] {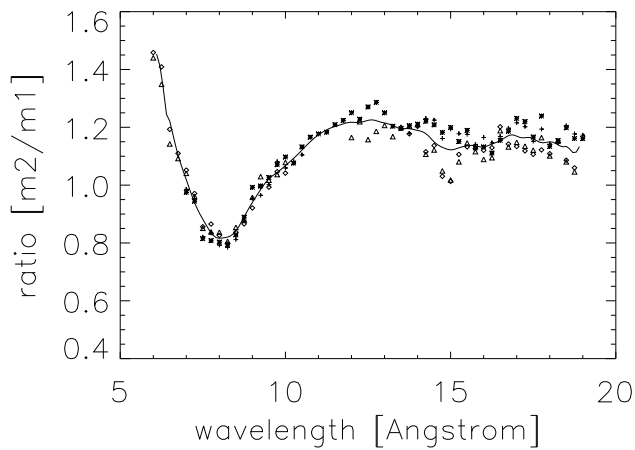}}
\caption{\small Correction factor for the m = --2 spectra relative to the m = --1 spectra (identical
for RGS1 and RGS2).}
\label{fig:science_m1m2}
\end{center}
\end{figure}

The {\it second order} effective area is normalized using the first order spectra, because the main emphasis of the
ground calibrations has been the first order only. The effect of this correction,
which is similar for RGS1 and RGS2, is typically of the order of
20 \% except at very short wavelengths (see Fig.~\ref{fig:science_m1m2})
and is well within the uncertainties of our theoretical model.

\subsection {Instrumental absorption edges}

{\it Oxygen edge:}
\label{sec:oxygen}
The instrument turned out to have a significantly larger instrumental
oxygen edge in orbit than anticipated. Some of this is due to oxygen contamination of the optics
(see Fig.~\ref{fig:RGA_response} for example), but a major contribution is on the detector itself. Following
launch the detector has always been cooled: before the detector
was opened its temperature was --50$^\circ$C  while following the opening of the 
detector the temperature dropped to --80$^\circ$ and subsequently around orbits 532--537 it was cooled to --110$^\circ$C.  In 
practice this means that any contaminating water ice, present on the detector body during launch, would
have remained on the detector and has never evaporated.

\begin{figure}
\resizebox{\hsize}{!}{\includegraphics[angle=0, clip, bb =60 360 540 630]{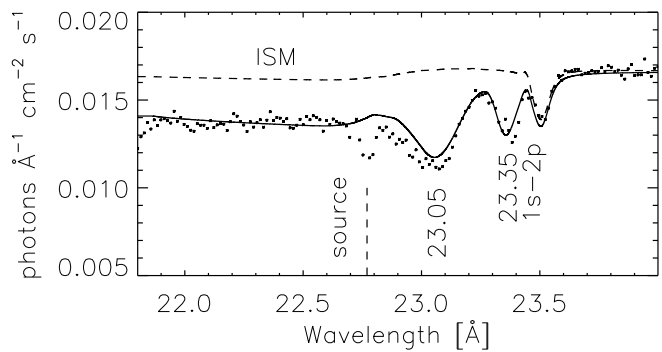}}
\caption{Combined spectrum around the O-edge for Mrk 421 and PKS 2155-304 with the
expected absorption from the ISM alone (dashed line) and in combination with the
the instrumental component (solid). The feature at 22.7 \AA~{} is not visible in other sources and
is therefore of astrophysical origin.}
\label{fig:oedge}
\end{figure}

\begin{figure}
\resizebox{\hsize}{!}{\includegraphics[angle=90, clip, bb =10 200 375 720]{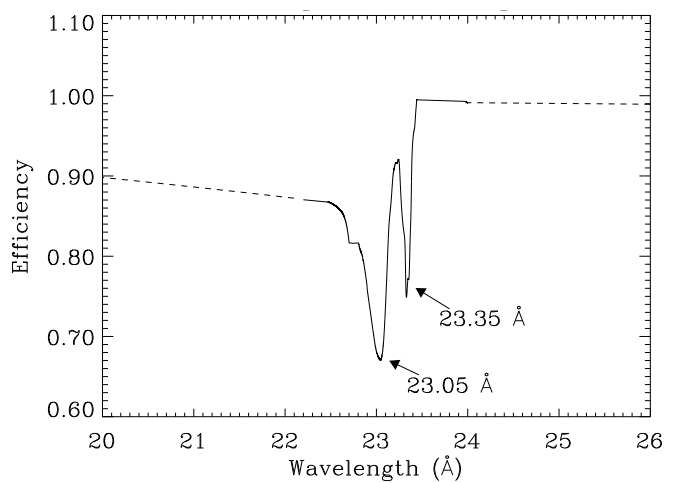}}
\caption{Instrumental efficiency around the O-edge}
\label{fig:instrum}
\end{figure}

In orbit the oxygen contamination has  been determined 
using relatively smooth continuum spectra. 
The oxygen absorption feature is illustrated in Fig.~\ref{fig:oedge}
where the spectrum for the combined data of two BL Lac objects with low extinction
are shown (172 ks for Mrk 421 and 346 ks for PKS 2155-304) as well as the expected
interstellar medium (ISM) extinction. There is a clear mismatch. Comparing these data with
sources with high extinction (Sco X-1) it is possible to deduce the instrumental
response over the O-edge. The result is shown in Fig.~\ref{fig:instrum}. See
\cite{devries2003} for details.
In the observed data the sharp features in this figure will be convolved with the broader instrument response.
Applying this instrumental correction to the low extinction data
improves the description considerably (see Fig.~\ref{fig:oedge}).
The feature at 22.77~\AA{} visible in this figure is not present in a
number of other spectra (Sco X-1, 4U 0614+19) and is therefore assumed to have an astrophysical origin.

\vskip 0.1 truecm

\noindent {\it Fluor and magnesium edges:}
On the CCD detectors there is a MgF$_2$ layer, which electrically isolates the optical Al filter from the CCD.
This layer will give rise to Mg and F absorption edges in the spectrum. By taking combinations of deep spectra
of continuum power law sources (Mrk~421, PKS~2155-304) and assuming no astrophysical features were present
around the Mg and F edges, the details of these edges were measured.
The Mg edge follows with sufficient accuracy the edge predicted by \cite{Henke}, but
the F edge shows deviations. The depth of the edge agrees with the depth predicted by Henke and the known thickness
of the MgF$_2$ layer, but the shape of the edge is different. This fitted shape is included in the effective 
area calibrations used by the SAS. 
 
\vskip 0.1 truecm

\subsection {Hot pixels}

\begin{figure}
\resizebox{\hsize}{!}{\includegraphics[angle=0, clip]{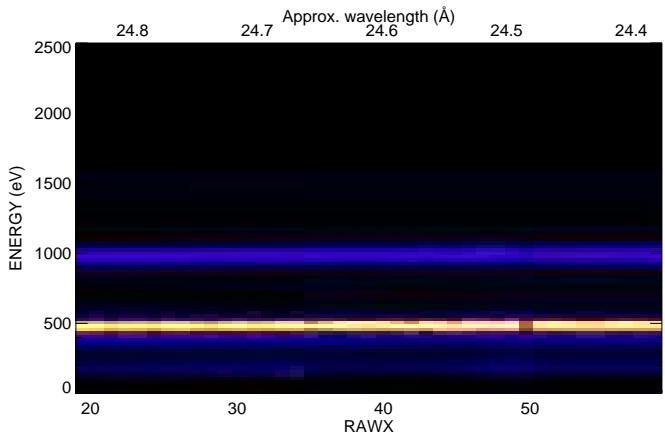}}
\caption{Signature of a cool column on CCD4 of RGS1 (1 Ms Mrk~421 data summed in detector space in a plot of 
pulse-height (Energy)
versus dispersion axis. Top curve is the second order of the dispersed spectrum, while bottom curve shows
first order. A "cool" column can be recognized around RAWX=50, or about 24.5~\AA{} in first order. This 
column extends into second order at the same position, but at half the wavelength. Top scale is the
approximate wavelength of the first order only.}
\vspace{-0.5 cm}
\label{fig:cool}
\end{figure}

In hot pixels the charge sometimes tends to overflow along the column to neighboring pixels. When pixels are
only occasionally hot during an observation, these pixels are not recognized as such, but the set of neighboring
pixels in a frame will be recognized by the onboard processing, and discarded as being a 'complicated' event or
cosmic ray. Such discarded pixels are not individually flagged, but only the average decrease in effective 
area of the whole CCD is recorded. Hence holes, or 'cool' pixels, appear in the processed spectra
as narrow 'absorption' features, due to the unrecognized loss of effective area for some columns. 
A typical 'cool' column is shown in Fig.~\ref{fig:cool}.

When searching for weak absorption features, it is important that these effects are taken into account. A list of 
pixels and columns is maintained in the calibration files (the 'COOLPIX' ccf). 
In the ground processing software (SAS) there is an option
to include or exclude these known pixels.
For individual observations, redundancy of both RGS instruments may allows us to identify
these columns in the individual instruments.
In addition spectra of first and second order are compared to reject absorption
features which occur at the same spot on the CCD, which is at half the wavelength in second order.
Furthermore the multiple pointing mode (section \ref{sec:multiple}) helps to mitigate this problem.

\subsection {Contamination}

Prior to launch the camera was purged with clean N$_2$ and its door was sealed. Between this
sealing and the launch many months passed. In this period water vapor may have entered
the cavity of the CCD detectors, especially during the time on the launch platform in the
very warm and humid air at the tropical launch site of Kourou, French Guiana.  

Following the launch the instrument was passively cooled to --50$^\circ$C and stayed
at this temperature until the opening of the detector door. Hence it is not surprising that a 
considerable oxygen edge is observed in the instrument response (see section \ref{sec:oxygen}).
Although the instrument has the capability
to be heated up to around 45~$^\circ$C this has not been exercised as it is virtually impossible to
predict accurately what happens to the instrument which has stayed cool for over 14 years. In addition
any contaminants, when evaporated, will largely stay inside the telescope tube and hence will condense again at a
later stage.

\begin{figure}
\resizebox{\hsize}{!}{\includegraphics[angle=90,clip]{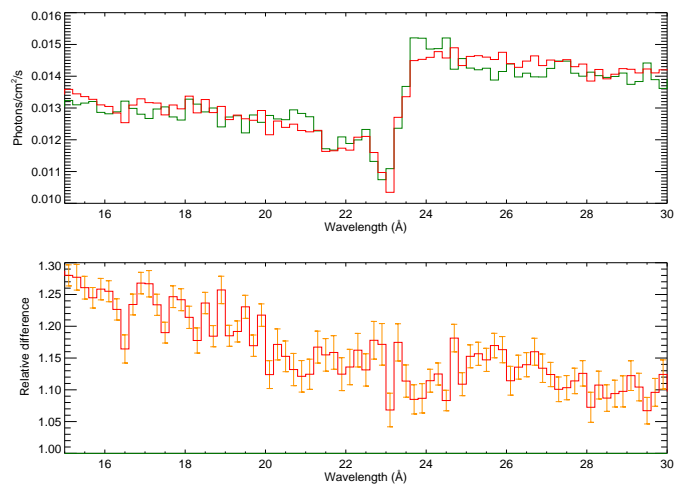}}
\caption{Stability of the instrumental oxygen edge: comparison of spectra corrected for a change
 in slope and normalization (top) and ratio (bottom) of Mrk 421 data  in orbit 84 (red) and orbit 2002 (green).
The spectra show the uncorrected instrumental edge, which is normally corrected for
during default data processing. The slight linear trend in the relative difference plot(bottom) is caused by small
changes in the source power law spectrum over time.}
\label{fig:oxygen_stability}
\end{figure}

The depth of the oxygen edge is very stable over time (see Fig. \ref{fig:oxygen_stability}).
Apparently the contamination with water ice indeed was a single event, and a single calibration (section \ref{sec:oxygen})
suffices to correct for this effect.

However, monitoring of two stable sources (RX~J1865.5-3754 and the Vela pulsar) shows that the response of the instrument
at long wavelengths suffers from loss of effective area, which increases over time. This is thought to be 
caused by a second contamination component: out-gassing of hydrocarbons by the carbon-fiber structures of the
telescope tube, although great care has been taken with selection of the materials. A special
cold section at --80~$^\circ$C is present in the telescope tube to freeze contaminants, and an out-gassing vent port
is present to release any remaining contaminants to open space. However contamination of the RGS CCD detectors
cannot be prevented because the CCDs are the coldest structures (--110~$^\circ$C) and completely open to the telescope tube.
By contrast, the EPIC cameras usually have optical blocking filters in front of them, which are at higher temperatures,
and protect their CCDs. 
In addition, contaminating layers may also form on the optics (mirror and gratings), which will have the same effect.

\begin{figure}
\resizebox{\hsize}{!}{\includegraphics[angle=90, clip, bb= 40 40 550 740]{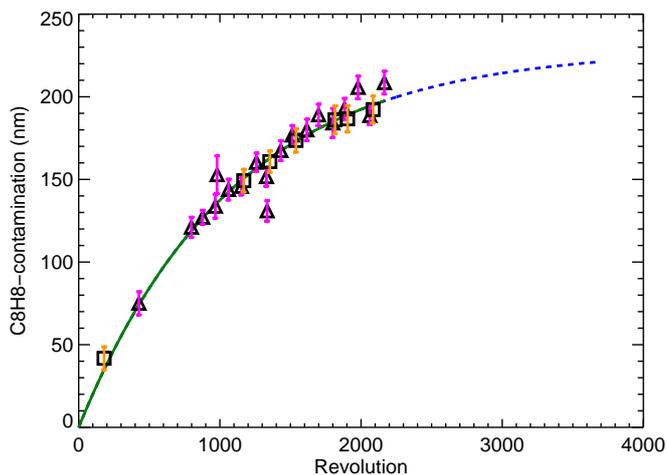}}
\caption{Contamination history given in units of C8H8 thickness, $\rho$ = 1.11 ${\rm g/cm}^3$.
Triangle data points are the RX J1856.5-3754 data points, while the square points are the
Vela pulsar data points. The broken line is the prediction of the contamination for the near
future.}
\label{fig:contamination-history}
\end{figure}

Assuming perfect stability of the calibration sources RX~J1865.5-3754 and the Vela pulsar at long wavelengths,
the thickness of the contaminating layer can be calibrated. Assuming a type of hydrocarbon as ${\rm C}_8{\rm H}_8$ to
be representative for the contaminating compound(s), Fig. \ref{fig:contamination-history} shows its measured thickness
with time.
After an initial increase of the carbon layer on the detector, the increase now levels off as the mission progresses.
For revolution 2200
a reduction of 47\% of the detector response at 35~\AA\ is observed. At shorter wavelengths, this reduction
is smaller e.g. 35\% at 30~\AA, and 24\% at 25~\AA. This reduction of effective area is taken
into account during standard processing.
 
\begin{figure}
\resizebox{\hsize}{!}{\includegraphics[angle=90, clip]{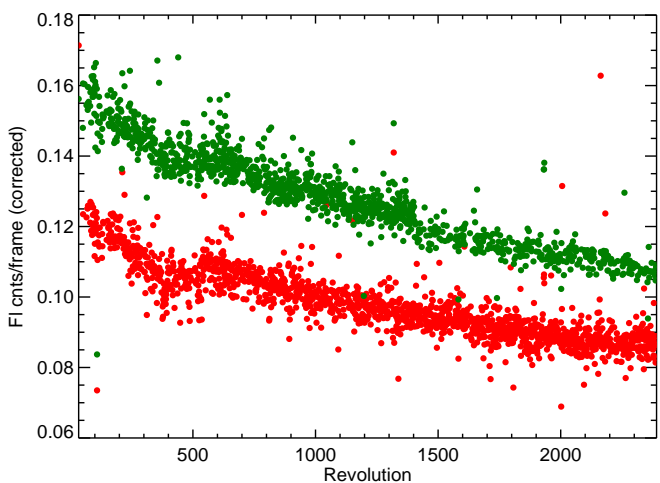}}
\caption{Onboard calibration source count rate for the Fluor source after correction for 
source decay time and CCD hot pixels and columns. Red data points is RGS1, green data points is RGS2.}
\label{fig:calcountsF}
\end{figure}

\begin{figure}
\resizebox{\hsize}{!}{\includegraphics[angle=90, clip]{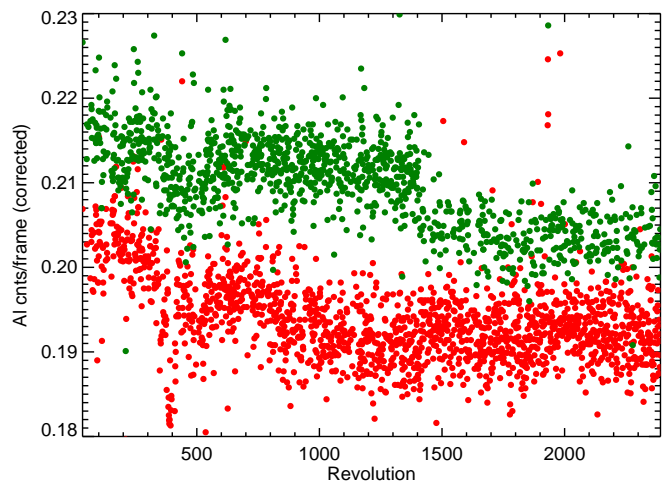}}
\caption{Onboard calibration source count rate for the Al source after correction for 
source decay time and CCD hot pixels and columns. Red data points is RGS1, green data points is RGS2. The drop in RGS2 count rate
around orbit 1405 is due to the switch to single node readout. A smaller drop at the same time is also observed for the fluor source
(Fig. \ref{fig:calcountsF}) }
\label{fig:calcountsAl}
\end{figure}

Using the onboard calibration sources in principle the contamination
of the CCDs can be verified independently. This is however complicated because the calibration source,
a Cm$^{244}$ $\alpha$ emitting source, is also cool and contaminated. Hence
contamination of the source or of the target will effectively lower this count rate more than is expected based on the
radioactive decay time of the source itself. Especially the alpha particles of the radio-active source, which are generating the
X-ray fluorescent photons on the F and Al targets, are very susceptible to small amounts of contamination. 
This is illustrated in Figs. ~\ref{fig:calcountsF} and ~\ref{fig:calcountsAl}. 
In addition, the hydrocarbon contamination will mainly affect the longest wavelengths but the Fluor-K line
and the Al-K line are at shorter wavelength (18.30 and 8.33~\AA{}, respectively) and hence far less
sensitive to contamination than the astrophysical calibration sources at 32 to 37~\AA.

\subsection{Pile-up effects}

Pile-up, which is the process of two X-ray events hitting the same or neighbouring pixel(s) within one readout frame, has the effect
on the effective area of moving two events from e.g. first order into one event in second order. Consequently
the effective area of the first order will decrease, while considerable background is added to second order.
Fortunately pile-up is negligible for most sources. 

Taking observations using different CCD integration times,
by using different total number of CCDs in the readout chain, on the same constant source, pile-up can be
measured. This way it is estimated that pile-up for the very bright source of the Crab is maximally 14\%
around 12~\AA{} wavelength \citep{Kaastra2009}, 
when using all CCDs in the default spectroscopy double node readout mode. However, Crab is an extended source. 
For the bright point source Mrk~421 in a flux state of 450 photons ${\rm m}^{-2}{\rm s}^{-1}{\rm \AA}^{-1}$ at 
$\lambda=25$~\AA, maximum pileup is estimated at
about 1\% for $\lambda=$16-18~\AA for double node readout. For RGS2 in single node
readout, pile-up is worse at about 2.4\%.
Most sources are much weaker than Mrk~421, and hence pile-up can be ignored. 

For very bright sources (e.g. Sco~X-1, Cyg~X-2) however pile-up can have dramatic effects. When there is considerable
pile-up, neighbouring pixels can combine and form 'complex' patterns, which will be discarded onboard.
Hence, holes will appear in the spectrum. There is no way to correct for those. Best is to limit pile-up by
using the RGS small window mode (see section \ref{sec:modes}), and/or using only a limited number of CCDs in the
readout chain.

\subsection {Effective area accuracy}

The absolute scale of the RGS effective area was calibrated on the well known source of the Crab pulsar and
nebula. This calibration has been described in detail by \cite{Kaastra2009}. It is estimated that the absolute
effective area scale of the RGS is accurate to within the accuracy of the adopted continuum model for the
Crab, which is about 10\%.

Contrary to most sources, the Crab is an extended source. This has been taken into account into the
effective area calibrations, which means that the instrumental models for off axis effective area and 
cross dispersion scattering were incorporated in the effective area analysis. In addition custom selections
had to be applied in the pulse-height versus dispersion coordinate space to prevent inclusion of too much
CCD background at long wavelengths and low pulse-height values. Details of this analysis are given in \cite{Kaastra2009}.
For on-axis point sources
these effects do not play a role. Crab is also a bright source, which means that pile-up has an effect.
As mentioned before, this effect was taken into account.

The redundancy between RGS1 and RGS2 can be used to obtain some idea about small scale systematic uncertainties
in the effective area, which define the sensitivity for detecting weak absorption structures. 
We start with the fluxed spectra ${\rm S}_\lambda$, in which all known exposure, bad columns and
pointing history effects have been taken into account in the used response. The $\lambda$ bins of the spectrum are
sufficiently narrow that they are of the same order as the wavelength range of individual CCD pixels, such that 
uncalibrated individual pixel effects will show up the the fluxed spectrum.
We know that on larger
scales ($>1$~\AA) residual uncalibrated differences exist in effective area between
corresponding bins of RGS1 and RGS2 and hence between the computed fluxes in the spectra. Small
scale spectral structures will be retained in the difference $\Delta {\rm S}_\lambda$ between the original spectrum 
${\rm S}_\lambda$ and the 1~\AA\ smoothed (convolved) spectrum ${\rm C}_{1\AA}({\rm S}_\lambda)$:
\begin{equation}
	\Delta {\rm S}_\lambda \equiv {\rm S}_\lambda - {\rm C}_{1\AA}({\rm S}_\lambda)
\end{equation} 
In effect this is a high spatial frequency filtering of the spectrum, which keeps the narrow features, by subtracting the smoothed spectrum in which all features have been smoothed over a sufficiently (1~\AA) wide area. 
For a featureless spectrum, the $\Delta {\rm S}_\lambda$ should follow the noise statistics of the spectrum, such that
\begin{equation}
	\sigma^2 = \frac{\sum{(\Delta {\rm S}_\lambda-\overline{\Delta {\rm S}_\lambda})^2}}{N}
\end{equation}
where $\overline{\Delta {\rm S}_\lambda}$ is the average value for $\Delta {\rm S}_\lambda$, which should be around zero,
$\sigma^2$ the variance of the distribution, and $N$ the number of spectral points. For the individual spectral
points ${\rm S}_\lambda$ the statistical error $\sigma_{\rm s}$ is known from
the photon statistics of the observation. For a featureless spectrum the average error should be identical to the computed
sigma: $\sigma = \overline{\sigma_{\rm s}}$, unless systematic effects play a role.

However, we do not know if the spectrum is really featureless. To get an idea of systematics we look at the difference
between RGS1 and RGS2. The individual spectra should follow the same true features, and hence these features are
eliminated from the difference:
\begin{equation}
	\Delta{\rm S}_{\lambda_{12}} \equiv \Delta {\rm S}_{\lambda_{\rm RGS1}} - \Delta {\rm S}_{\lambda_{\rm RGS2}}
\end{equation}
The variance of this subtraction ($\sigma_{12}^2$) is twice the variance of the individual $\Delta {\rm S}_\lambda$:
\begin{equation}
	{\sigma_{12}}^2 = \frac{\sum{(\Delta {\rm S}_{\lambda_{12}}-\overline{\Delta {\rm S}_{\lambda_{12}}})^2}}{N} = 2 \cdot \sigma^2
\end{equation}
Table \ref{tab:r1r2dif} shows $\sigma$ computed from the RGS1-RGS2 difference, expressed in units of error on
the individual spectra points $\sigma_{\rm s}$
for the observed flux in Mrk~421, by combining different sets of observations to obtain different signal
to noise ratios of the spectrum.
%
It appears that, when signal to noise increases the $\sigma$ of the distribution
becomes larger than the statistical error $\sigma_{\rm s}$ because
systematic uncertainty effects, the pixel-to-pixel systematic errors, become more dominant. This means
that a larger fraction of pixels will have values outside the statistical expected accuracy (the outliers). This
effect should be taking into account when looking for weak line features close to the statistically expected threshold.
For signal to noise ratios larger than 10 (100 counts per bin) these effects start to play a role. For signal to noise
ratios of around 70, the statistical uncertainty is equal to the systematic uncertainty, translating into about a 1.5\% 
uncertainty in short scale effective area variations.  

\begin{table}
\caption{Differences between RGS1 and RGS2 as function of signal to noise ratio
per pixel. Differences outside normal Poisson statistics reflect the systematic uncertainties
in the effective area accuracy}
\label{tab:r1r2dif}
\centering
\begin{tabular}{ccc}
S/N ratio & difference \ann{a} & outliers \ann{b} \\
          & width ($\sigma_{\rm s}$)   & (\%)   \\ \hline
6.7	&	1.01	&  0.2	\\
36	&	1.09	&  0.3  \\
69	&	1.41	&  3.7  \\
\end{tabular}
\begin{minipage}{7cm}
\begin{itemize}
\setlength{\itemsep}{0pt}
\item[\ann{a}] Width of distribution of differences expressed in statistical uncertainty $\sigma_{\rm s}$ of the spectrum. Offsets
(constant difference) are ignored.
\item[\ann{b}] Percentage of pixels outside of the $3 \sigma_{\rm s}$ statistical uncertainty.
\end{itemize}
\end{minipage}
\end{table}


\section{Instrumental background}

The instrumental background consists of several components, each with their own
characteristics and temporal behavior.
In the next subsections we will discuss the characteristics and behavior of these components, 
and the way background can be dealt with in the observational data.

\subsection{Soft proton background}
\label{sec:softp}

\begin{figure}[ht] 
\resizebox{\hsize}{!}{\includegraphics[clip]{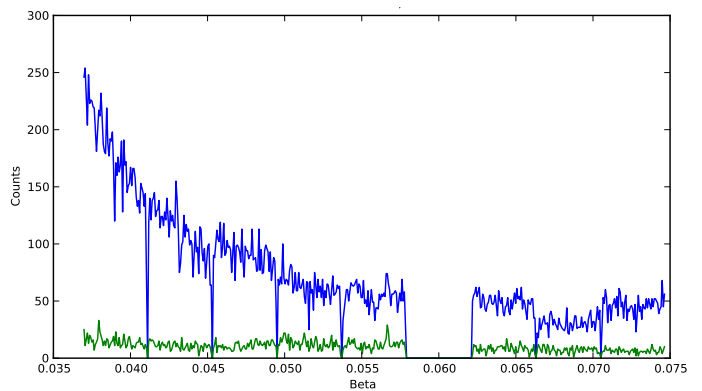}}
\caption{Raw count rate distribution as function of CCD position along the dispersion direction for an off-axis 
region. The difference between a quiet period (green line) and a high background period (blue line) is 
in agreement with the expectations. The observations used are from AB Dor, RGS2, revolution 546 (high background) and
revolution 668 (low background), with similar cross-dispersion and pulse-height selections for all CCDs.}
\label{fig:proton}
\end{figure}

The major time dependent component of the instrumental background are the soft protons, trapped in the
earth magnetic field and scattered through the XMM optical path.
Apart from the potential of causing damage on the CCDs (see sections \ref{sec:protscat} and \ref{sec:ccdrad}),
soft protons mainly add a time variable background component on the detectors.
Due to the scattering characteristics of the mirrors and gratings, soft protons will have the highest flux closest
to the optical axis, on CCD9, decreasing towards CCD1. Selecting data cuts outside the parameter space of
the normal X-ray dispersed spectrum on CCD9 provides a high quality monitor on the soft proton flux and the
nature of the main instrumental background component. In orbit, the count rate on RGS CCD9 is used as a radiation 
monitor to warn for high incident proton flux, and is used to activate the protection for the EPIC instruments.
Fig. \ref{fig:proton} shows the nature of the soft proton flux as counts on the off-source positions, integrated
over the full energy range as function of the dispersion axis, comparing a period of high soft proton activity 
with a period of low activity.

To improve the signal to noise ratio of the processed spectra, the count rate on CCD9 can be used to select periods of
low soft proton background. For extended sources, count rates on CCD9 outside the normal spectra pulse height
space 
are used as parameters to construct a background spectrum, based on spectra
of empty fields with similar CCD9 count rate (soft proton) parameters. These sets of background spectra for various 
soft proton backgrounds are maintained as standard calibration files for the RGS processing software (SAS). 

\subsection{High energy particle background}

Soft protons are low energy particles, scattered by the optical components. High energy particles, mainly from
solar CME events, simply penetrate through all spacecraft structures, evenly illuminating the RGS CCD detectors.
High flux events only occur occasionally and are usually predicted, in which case the instruments are switched off.
Those events do however increase the amount of hot pixels and columns and CTI of the CCDs.
At all times there is a low flux, high energy particle background. Many of these particles are detected
and discarded as events well above the upper energy threshold due to their high energy contents, or are 
recorded on the CCDs as a track across many pixels, corresponding to the particle path across the CCD, and are discarded by the 
onboard processing as 'complicated' events. The average effective area of the CCD is corrected for the
pixels discarded in this way. It is thereby assumed that this background does illuminate the CCD evenly.
The remainder of these events, which are not recognized as such add a low and flat background component.

\subsection{Detector background}

The CCD detectors produce low energy system noise and warm/hot pixels. These effects have been
described before (sections \ref{sec:ccdnoise} and \ref{sec:defects}). Applying the appropriate selections
in data space and on pixel level, this background is minimized.

The calibration sources are a type of background introduced on purpose on the detector. This background is
very constant and well known. The main part of this background is discarded by the
appropriate selections in data space. The remaining part is
perfectly modeled by the standard processing tools.

Another type of detector background is due to fluorescent lines generated by the high energy particles
background in the detector housing. This background is variable. Fluorescent lines that can
be recognized are the Al-K line because most of the material of the housing is aluminium and the gold-M line
 because there is a thin layer of gold deposited on the structures surrounding the CCDs. These lines can be
recognized in CCD pulse height space only and are constant in dispersion and cross-dispersion directions. 
Hence they do not show structure in the dispersed X-ray spectrum, but do add e.g. 0.02 cnts/s over
the entire CCD detector for the gold-M line.
 
\subsection{X-ray background}

The sky itself provides a true X-ray background, consisting of numerous unresolved distant cosmic sources
\citep[see e.g.][]{gili2007},
and X-ray lines generated locally by e.g. the charge exchange mechanism in the earth vicinity in interplanetary space
\citep[see e.g.][]{cox1998,cravens2000,fujimoto2007}.
For the RGS the X-ray background can be treated as an extended source. X-ray lines from the background will be visible in the
RGS spectra as very broad features which follow the effective area as function of dispersion offset. 
With standard background processing for point sources, which takes the background from off-axis
positions in cross dispersion directions and applying the same selections in CCD pulse height space, 
the X-ray background is perfectly subtracted, since it is identical for source and off-source positions.
When applying background models for extended source analysis, discrepancies between X-ray backgrounds from the sky positions from
which the models are made with the observed sky position can keep some background signal in the spectrum. 
That signal however is usually very low. 


\section{Instrument operations}

Instrument operations  had a clear impact on the calibrations. Onboard hardware settings
affect the efficiency with which the charge can be clocked through the CCD, the selected 
read-out speed affects the amount of pile-up, onboard event selection parameters  
influence the number of selected events to be transmitted to the ground and so on.
In this section we present the operational aspects of the instrument that 
affected the calibrations. Table \ref{tab:events} lists the major changes in instrument
operations. We first start with the operational conditions that affected the hardware
configuration of the instrument.
\label{sec:operations}

\subsection{Instrument operational conditions}

{\it Detector element failures}: the read-out channels connected to two CCDs failed over the mission lifetime.
The first failure (RGS2 CCD4) was traced back to a frequently used digital hardware switch (HS-202RH) whereas
the second failure (RGS1 CCD7) could not be unambiguously traced back to a single hardware component and 
could also be due to failure in the CCD itself.
Due to the complete redundancy between the
two instruments, the full wavelength range is still covered. In particular around the \ion{Ne}{x} Ly$\alpha$
line and the \ion{O}{vii} triplet this has significant impact.

After the second failure it was decided to lower the clock voltages to reduce the load on various electrical components 
(orbit 168, see Table \ref{tab:events}). 
This tuning resulted in a slightly modified instrument response which can be completely accounted for in the calibrations.
\vskip 0.3 cm

{\it Perigee passage condition}: during the initial phase the instrument was powered during the perigee passage. From
orbit 171 this powering during perigee passage was stopped because there
was concern about the de-rating of some of the electronic components related to the above mentioned failures. Because the
instrument behaved stable over many years after this change, this concern was probably not justified. 

\vskip 0.3 cm

{\it Detector biasing}: Radiation damage at the CCD gate structure translates into a different biasing of the
clocks used to transfer the charge through the CCD and thus results into a degradation of the charge transfer efficiency (CTE). 
By the capability of the instrument to set the clocking voltages in each CCD such degradation can be compensated.
Apart from re-tuning to set proper clock voltages following the clock voltage reduction after the RGS1 CCD7 failure,
changes in clock voltage settings have only been necessary for one CCD. The clock voltage of RGS2 CCD2 was modified in
orbit 1389 (see Table  \ref{tab:events}).

\vskip 0.1 cm

{\it Cooling}: Initially the detectors were operated at --80\celsius. This temperature was selected as a compromise 
between the charge transfer inefficiency (which gets better for temperatures as low as --110\celsius\, or --120\celsius) 
and the potential contamination on the detector. Because there is no thermal filter between the detector surface and the 
{\it XMM-Newton} telescope tube, the structural element connecting the mirror platform and the detector platform, 
the camera will act as cryo pump for this volume. The relatively high operational temperature resulted in an increasing 
number of hot and flickering pixels during the first years in orbit. In orbit 532 and 537  the operating temperatures 
of the detectors were lowered to --110\celsius\, to reduce the fraction of hot pixels.

\vskip 0.1 cm

{\it Current limiter RGS2}: Each CCD is read out simultaneously through two output amplifiers and only one CCD is 
read out at a time: while one CCD is read out, the others accumulate data. Upon completion of
the read-out, the image section of the next CCD is transferred to the storage section and this next CCD is ready 
to be read out.

To handle the data streams each camera has two complete read-out chains, which control the clocking of the charge
through the CCD, provide the analog to digital conversion and perform data selections to keep the telemetry within
reasonable limits. Each read-out chain includes circuits to protect the rest of the system for potential failures.
The dimensioning of one of the current limiters in one of the read-out chains (RGS2, node D) was too restrictive to 
cope with the de-rating of its electrical components beyond the planned lifetime. This happened after about 4 years. Initially the
camera continued to operate properly as the current limiter was reset by the 'power-on' of the
CCDs following tripping of the current limiter. Due to
the continued degradation of the components, the occurrence of this problem increased and started to affect
the measurements; there was a need for more frequent resets. 
By lowering the temperature of the electronic components ($<$ 4$^\circ$C) the current limiter was pushed back 
into its operating range. Further degradation of the components required an even lower temperature of the 
electronics after orbit 1408 (August 27, 2007). This could only be achieved by lowering the temperature of 
the full focal plane and this would affect the calibrations of the other instruments on {\it XMM-Newton} too much. 
Therefore it was agreed to disable one of the two read-out chains of RGS2 and read its charge through a single 
output node by changing the clock patterns. The only disadvantage is that this increases the read-out time of all 
CCDs by approximately a factor 2 resulting in a slightly larger dark current and a larger fraction of piled-up 
events. For most of the observations, however, the pile-up fraction is very small and therefore the impact is 
marginal. An unexpected result of this change in read-out
was that the fixed pattern noise (see section \ref{sec:ccdrad}) disappeared.

\begin{table}
\caption{Major changes in RGS hardware operations and their scientific impact. At this moment, 
after 2009 no major changes have been necessary.}
\label{tab:events}
\begin{center}
\begin{tabular}{rrp{5.5cm}}
\hline
orbit	& date  	&	event \\
	& yy-mm-dd	&	\\
\hline
24 	& 00-01-25 	& first light RGS2 \\
26	& 00-01-30 	& failure RGS2 CCD4 electronics \\
	&		& \mbox{$19.9<\lambda<24.9$~\AA} \\
30	& 00-02-06 	& first light RGS1 \\
45	& 00-03-08	& upload DPP code (19) for improved onboard particle background data handling. Data $<$  orbit 45 have wrong exposure \\
52	& 00-03-21	& update IC code (S206): no impact \\
135	& 00-09-02	& failure of RGS1 CCD7 \\
	&		& \mbox{($10.3<\lambda<13.6$~\AA)}, RGS operations suspended \\
147	& 00-09-26	& operations resumed \\
168	& 00-11-07	& modification of bias settings of RGS to reduce part stress \\
171     & 00-11-13      & RGS set to setup mode during perigee \\
192	& 00-12-25	& RGS2:CCD2 bias back to before orbit 168 \\
361	& 01-11-29	& upload IC code (S207): no impact \\
363	& 01-12-01	& upload IC code (S207): no impact \\
532	& 02-11-04 	& cooling RGS2 to -115$^\circ$C  \\
537	& 02-11-13 	& cooling RGS1 and RGS2 to -110$^\circ$C  \\
538	& 02-11-15	& RGS2 CCD2: serial bias to -2.7 V \\
584	& 03-02-15	& CSG code (83) : HTR 3x72 mode \\
765	& 04-02-13	& 1st activation of current limiter RGS2\\
786 	& 04-03-24	& 2 hot spots in RGS1:CCD1 \\
905	& 04-11-18	& RGS1 upload IC code (S208): no impact \\
907 	& 04-11-22	& RGS2 upload IC code (S208): no impact \\
1207 	& 06-07-12	& lowering temperature electronics to limit impact of current limiter \\
1389 	& 07-07-11	& RGS1 CCD2: serial bias to -4.8 V \\
1408 	& 07-08-16    	& single node read-out RGS2 \\
1671 	& 09-01-24	& upload CSG code (95): small window mode became available to users after 10-10-8 \\
\hline
\end{tabular}
\end{center}
\end{table}

\subsection{Operational settings}

\vskip 0.1 truecm

In addition to these changes in the hardware configuration, there is a large set of onboard parameters
and operating modes which may effect calibrations and/or data quality:

{\it Lower level threshold}: The unprocessed data stream from the CCDs is too large to be handled by
the onboard electronics. Therefore we reject all bins with a contents consistent with the 
CCD system level plus dark current and (read-out) noise. This is to so-called rejection threshold.

{\it Hot pixel and column rejection}: In addition the pixel coordinates are
compared to an uploaded hot pixel/hot column table and are rejected if they are listed in this table.

{\it Event recognition}: The remaining data are searched for valid X-ray patterns. Events
which spread their charge over an area which extends beyond a 2x2 pixel square are rejected 
(charged particles or piled-up events). After adding the charge for the
remaining valid patterns, only reconstructed events are transmitted to the ground if their charge is 
above acceptance threshold and below upper threshold. The acceptance thresholds are set such that they approximately 
follow the lower boundary of the pulse-height distributions of first order dispersed spectral events. 
These thresholds are set per half of a CCD (CCD-node), and explain
the increase of dark areas towards low $\beta$ angle in the bottom panel of Fig. \ref{fig:data_space}.

{\it Instrument modes}: The read-out of each CCD can be controlled in great detail. For the RGS operations
it turned out that $ 3 \times 3 $ onboard binning of the pixels was a good compromise between the read-out speed,
the number of hot columns, and the line spread function. For high count rates one can reduce the effect of pile-up,
meaning the effect of more than  one photon in a bin during the data accumulation, by either reducing the number
of CCDs to be read out, or by selecting a smaller region in the cross dispersion direction. This later option,
the so-called "small window mode", is available from orbit 1671 onward.
\label{sec:modes}

{\it Multiple pointing}: Despite all calibration efforts it cannot be excluded that there are small temporal unaccounted for
CCD pixel to pixel variations in e.g. electric offset and/or charge loss due to evolving CCD damage.  
This varying CCD response effects the effective area by means of the event selections, and can affect
the interpretation and identification of weak emission or absorption
lines. By splitting long observations into observations with shorter observation time and using different pointing offsets
in steps of 35 m\AA{} corresponding to steps of 15 arcsec, 
the detector response can be separated from real features in the observed source. Although
this was feasible from the launch onward, from "call for proposals" AO08, orbit 1693, the so-called multi-pointing mode
has been made available to
general users. There is no overhead during re-pointing of the telescope, since all data, including the data obtained during 
the small slews of the telescope, can be used. Locations of X-ray events on the detector are converted
into wavelength space, based on 
telescope pointing reconstruction at the moment of X-ray detection. 
\label{sec:multiple} 

\subsection{Radiation environment}

\begin{figure}
\resizebox{\hsize}{!}{\includegraphics{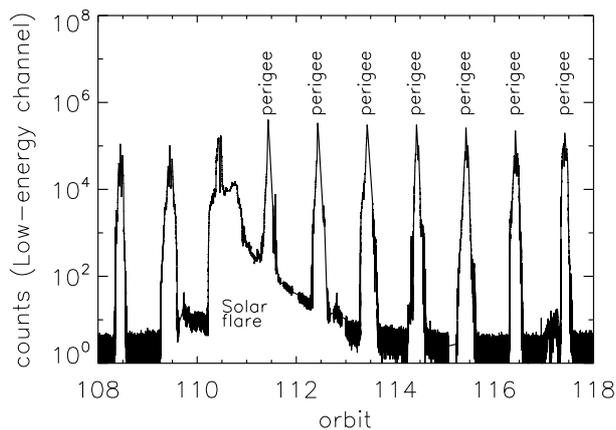}}
\caption{Radiation environment for \xmm, as observed by the radiation monitor in orbits 105-120, 
including 10 perigee passages (indicated) and one of the strongest CME observed.}
\label{fig:solar}
\end{figure}

The radiation environment is characterized by three distinct components: the orbital passage through the 
radiation belts, coronal mass ejections (CMEs) due to solar activity and regions in the orbit with
trapped soft protons. In Fig. \ref{fig:solar} two components are shown: the regular transfer through the
radiation belts and the presence of one of the major CMEs during the mission lifetime.

\vskip 0.1 cm

{\it Perigee passage}:
Up to orbit 171 the instrument was kept switched on during the perigee passage and indeed the CCDs were completely
flooded by charged particles (as could be deduced from the full frame images). After this orbit the 
instrument was switched off (with the exception of the Digital Electronics) to minimize possible damage by energetic
charged particles of the different electronic components.

\begin{figure}
\resizebox{\hsize}{!}{\includegraphics[clip,angle=90]{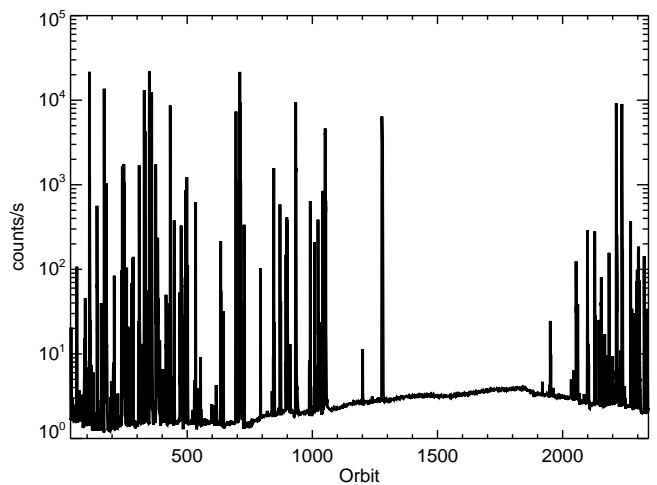}}
\caption{Count rate outside the radiation belts registered by the \xmm{} radiation monitor 
in the high energy "nhe2" channel (electrons: 1.3-1.75 MeV, protons: 9-37 MeV). 
This displays the solar activity (spiky structures due to solar flares) throughout
the mission, showing the active sun at the start and after 10 years of mission and the solar minimum in between. }
\label{fig:solaractiv}
\end{figure}

\vskip 0.1 cm
{\it Solar activity}: Fig. \ref{fig:solaractiv} shows all solar activity throughout the mission as recorded 
by the \xmm{} radiation monitor. In Table \ref{tab:solar_activity} we list the major flares. We provide the combined maximum count rate for electrons between 130 keV and 1.5 MeV and protons between 1 and 4.5 MeV during the increased solar activity. 
Clearly the strong flares last for several up to 15 revolutions. As is indicated in this table, \xmm{} has experienced 5 major solar flares with log(cnts/s) $\ge$ 4.
With use of
the OMNI data (see http://omniweb.gsfc.nasa.gov/ow.html) we have calculated the fluence and hardness ratios
of these flares. Due to the incompleteness of the data (telemetry) we can only provide lower limits 
for the fluences but these are reasonably representative as the number of missing data points is limited. 
For the first 4 major flares the fluences and hardness ratios are given in the table. 
All other flares are significantly weaker.
In section \ref{sec:ccdrad} we described the effect of these flares on the detector response.
 
\begin{table*}
\caption{Major solar activity, which affected the observations and/or operations}
\label{tab:solar_activity}
\begin{tabular}{rrcccl}
\hline
orbit	& date  	& XMM		& fluence \ann{a}	& hardness 	& comment/duration \\
	& yy-mm-dd 	& log[cnts/s] 	& ($\times 10^9$) & ratio \ann{b}	&		\\
\hline
110	& 00-07-15	&	4	& 1.2		& 0.48		& 3 revolutions \\
\hline
169	& 00-11-10	&	4	& 1.6		& 0.38		& 1st burst, 13 revolutions \\
176	& 00-11-25	&		&		&		& 2nd burst \\
\hline
239 &	01-03-30	&	3	&		&		& 9 revolutions \\
\hline
375	& 01-12-27	&	3	&		&		& 1st burst, 12 revolutions \\
383	& 02-01-11	&		&		&		& 2nd burst \\
\hline
433	& 02-04-21	&	3	&		&		& 5 revolutions \\
\hline
711	& 03-10-28	&	4	& 1.0		& 0.28		& 1st burst, 15 revolutions \\
723	& 03-11-20	&		&		&		& 2nd burst \\
\hline
934	& 05-01-15	&	4	& 0.8		& 0.25		& 5 revolutions \\
\hline
1052 	& 05-09-07	&	3 	&		&		& 5 revolutions \\
\hline
1282 	& 06-12-09	&	3 	&		&		& 6 revolutions \\
\hline
2106    & 11-06-09      &       3       &		&		& 2 revolutions \\
\hline
2220    & 12-01-22	&	3	&		&		& 4 revolutions \\
\hline
2242	& 12-03-06	&	4	&		&		& 2 revolutions \\
\hline
\end{tabular}
\vskip 0.2 cm
\begin{minipage}{10cm}
\begin{itemize}
\setlength{\itemsep}{0pt}
\item[\ann{a}] Protons above 2 MeV (lower limit).
\item[\ann{b}] p~($\geq$10 MeV) $/$ p~($\geq$ 2 MeV), with p number of protons.
\end{itemize}
\end{minipage}
\end{table*}

\begin{figure}
\resizebox{\hsize}{!}{\includegraphics[clip]{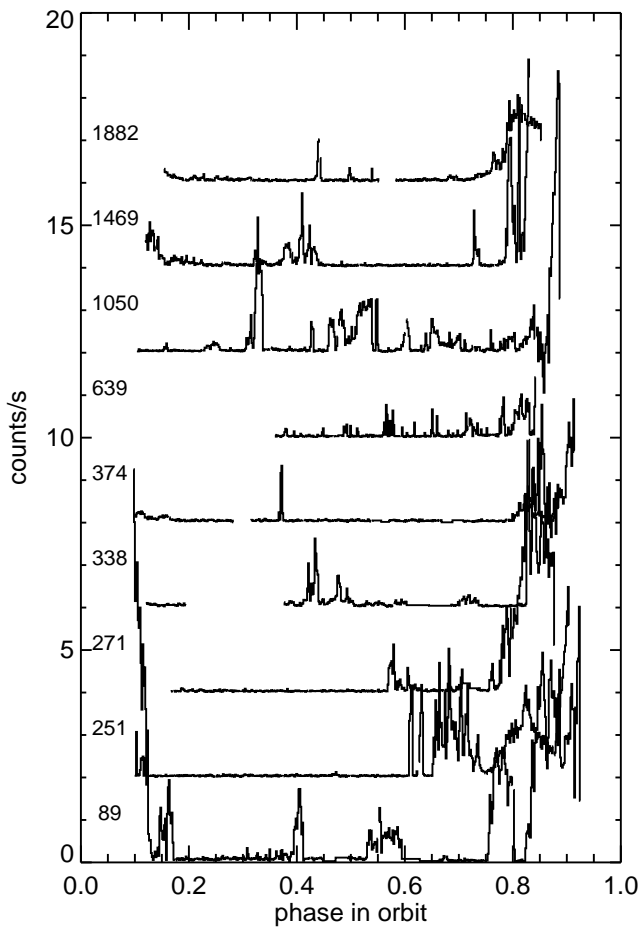}}
\caption{Illustration of soft proton count rates per second in the RGS detector (CCD9 node C, RGS1) for 9
different orbits (displayed with an offset of 2 for clarity)}
\label{fig:soft-protons}
\end{figure}

{\it Soft protons}:
In addition to these expected radiation components 'clouds' of soft protons are present in the environment.
Typically their occurrence increases closer to perigee but they can occur in short bursts all over
the orbit.
In Fig. \ref{fig:soft-protons} we show the typical count rates during a random set of orbits. The
count rates shown are data collected in the RGS instrument with pulse heights and positions on the detector where
no X-rays dispersed from the gratings are collected (regions outside the source position, on CCD 9). Soft protons
from outside will be absorbed by 
the satellite and detector housing but protons entering the telescope mirrors are scattered by the optics and subsequently
recorded by the detector (see
section \ref{sec:protscat}). Therefore this count rate is a reasonable accurate estimate of this background component.

\begin{figure}
\resizebox{\hsize}{!}{\includegraphics[clip,angle=90]{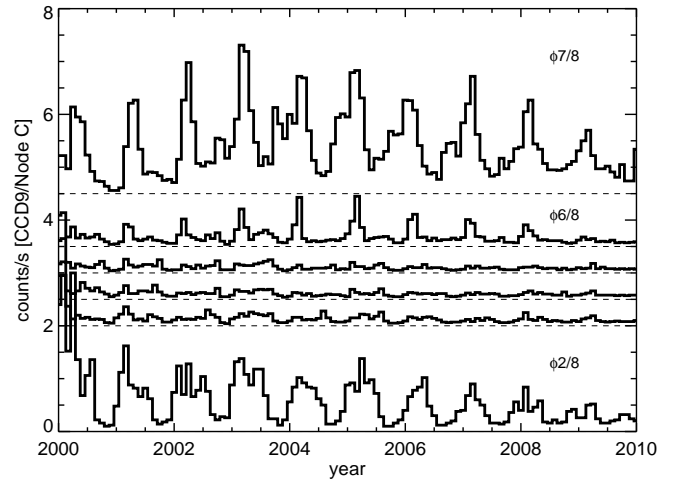}}
\caption{Seasonal variations in the instrument background in counts per second on node C of CCD9, as 
function of orbit. The different phases of the orbit are shown from bottom to top with a different
offset for clarity (indicated by the dashed lines). The phases for the perigee passage are omitted
(1/8 and 8/8). The others are shown for
increasing phase (bottom to top; for clarity the labels of phases 3/8 to 5/8 are left out). 
The large scale trend of increasingly smaller maxima in later years,
shows the decreasing solar activity over the 11 year solar cycle.}
\label{fig:seasonal}
\end{figure}

{\it Seasonal variations}: 
The background rates during passage through the perigee belts shows a clear dependency on the season. During the
course of the seasons the relative position of the orbit with respect to the Earth magnetosphere changes.
This is illustrated in fig.~\ref{fig:seasonal}
where the count rate in RGS for $\lambda  <  7.7$~\AA{} and a selection in the cross-dispersion direction
$> 1$ arcmin away from the source direction is shown for different
phases of the orbit. During the first and last part of the orbit the instruments are switched off since
during the perigee passing no science data can be recorded. In addition to reduced background closer to
apogee  we see also the seasonal effect of the impact of the 
Earth magnetosphere. 

\section{Conclusions}


Continuous monitoring and special purpose calibration observations of the RGS instruments over
the years, since launch, have yielded an accurate calibration and understanding of the instruments. Although many
instrument parameters were measured before launch, in flight measurements of celestial sources allowed
a far more accurate assessment of these parameters. In addition the behavior of the instrument due to 
cosmic radiation, temperature and attitude variations, and aging, are now well understood.

The following major aspects of the RGS, described in this paper, can be summarized as follows:    

\begin{itemize}

\item{{\it Detector performance} The CCD detectors are subject to damage by cosmic radiation and soft protons.
The major effect is the appearance of hot pixels and columns. Cooling the detectors to -110$^\circ$C made most hot pixels
disappear and has kept the amount of hot surface limited since then. Unfortunately two different CCDs were
lost in the two instruments quite early in the mission, but overlap between the instruments still allows the full spectrum to be
measured. Effects of increasing charge transfer inefficiency (CTI) are small ($< 10\times10^{-5}$) 
and appear not to play a role in the RGS.}   

\item{{\it Wavelength scale} Although the original calibration kept the wavelength scale accuracy within specification,
improvements have been made by taking into account effects of the angle of solar illumination on the
spacecraft and seasonal effects in pointing. The blind accuracy of the wavelength scale is now of order 6 m\AA{} in first order. }

\item{{\it Line spread function} The observed line shape accurately follows the model. The line center appears
slightly narrower than originally predicted by the models. The response has been updated for that.}

\item{{\it Effective area} The effective area is subject to bad pixels and columns and the two failed CCD chains 
(see Fig. ~\ref{fig:aeff}). In addition, a constant absorbing layer of water and an increasing column of 
hydrocarbons decreases the effective area. At this moment (revolution 2500, almost 14 years in orbit), the effect 
of the hydrocarbons is a loss of 47\% of effective area at 35 \AA{} and 24\% at 25 \AA. These effects are all taken
into account in the standard calibration. The absolute accuracy of the effective area is estimated at 10\%.}

\item{{\it Sensitivity} Deeper integrations in principle allow detections of weaker (absorption) lines, but do expose
systematic instrumental structures, which may be confused with real astronomical features. Systematic structures can
be identified by utilizing the intrinsic redundancies in the instrument (two RGSs, two orders). In addition, a multi
pointing mode has been made available which is highly effective in decreasing the systematic errors and helps in
identifying the remaining systematic structures. Ultimate sensitivity for weak absorption lines is reached when the
statistical uncertainty matches the systematic uncertainty. This condition is met for absorption lines with
depths of about 1.5\% of the continuum.}

\end{itemize}

The RGS instruments onboard \xmm{} are in excellent shape, and have well exceeded their design life time (5 years).
The actual performance has fulfilled the design goals. Detector (CCD) degradation due radiation and contamination
has been limited and extrapolations show that detector performance will be quite adequate for at least 
another 20 years. Continuing calibrations and ongoing analysis of the instrumental
behavior will keep this instrument in excellent shape in the years to come. 
 

\vskip 0.5 cm

\begin{acknowledgements}
The authors like to thank the operations teams at ESOC and ESAC for their full support. In addition
a number of colleagues have contributed significantly to the paper including in particular Ricardo Perez Martinez.
We also like to thank Jacco Vink for his work on the effective area corrections, and
Yan Grange for looking into extraction methods and background issues. Norbert Schartel gave some useful comments
on a first version of the text.
XMM-Newton is an ESA science mission with instruments and contributions directly funded by
ESA Member States and the USA (NASA). SRON is supported financially by
NWO, The Netherlands Organization for Scientific Research.
\end{acknowledgements}

\bibliographystyle{aa}
\bibliography{ref}


\onecolumn
\appendix
\section{Calibration files}
\vskip 0.5cm
\tablecaption{Current Calibration Files (details on CCF release are available on
http://xmm2.esac.esa.int/external/xmm\_sw\_cal/calib/rel\_notes/index.shtml)}
\tablefirsthead{
& orbit 	& validation		&	RGS		& RGS  		& comment \\
&	&	start date	& 1			&  	2	&	\\	
&	&	yy-mm-dd		& 			&		&  \\		
\hline
}
\tablehead{
& orbit 	& validation		&	RGS		& RGS  		& comment \\
& 	&	start date	& 1			&  	2	&	\\	
& 	&	yy-mm-dd		& 			&		&  \\		
\hline }
\begin{supertabular}{lllrrl}
\label{table:CCF}
{\bf ADUConv}
& 25	& 00-01-25		&			& 006		&  values for launch \\
& 30	& 00-02-06 		&   006 		& 007 		&  values for launch\\
& 105	& 00-07-09		&   007			& 008		&  regular update \\
& 114	& 00-07-23		&   008			& 009		&  regular update \\
& 139	& 00-09-12		&   009			& 010		&  regular update \\
& 169	& 00-11-11		&   010			& 011		&  regular update \\	
& 191	& 00-12-25		&			& 012		&  regular update \\	
& 259	& 01-05-10		&   016			& 016		&  regular update \\			
& 325	& 01-09-18 		&   017			& 017		&  regular update \\
& 349	& 01-11-05		&   018			& 018		&  regular update \\
& 532	& 02-11-04		&			& 020		&  RGS2 shortly at -115$^\circ$C \\
& 536	& 02-11-13 		&   019			& 021		&  update after cooling to -110$^\circ$C \\
& 808	& 04-05-09		&   020 		& 022		&  regular update	\\
& 1400	& 07-08-01		&   022			&		&  regular update \\
& 1408	& 07-08-17		&			& 025		&  RGS2 single mode \\
& 1839  & 10-01-01		&   023			& 026		&  regular update \\
& 2021	& 11-01-01		&   024			& 027		&  regular update \\
\hline
{\bf CTI}
& 1	& 00-01-01		&   002			& 002		&  values for launch \\
& 259	& 01-05-10		&   003			& 003		&  regular update \\
& 532	& 02-11-04		&			& 004		&  RGS2 shortly at -115$^\circ$C \\
& 536	& 02-11-13		&   005 		& 006		&  update after cooling to -110$^\circ$C \\
& 807	& 04-05-09		&   009 		& 009 		&  regular update  \\
& 1408	& 07-08-17		&			& 011		&  single mode read-out \\
& 1839  & 10-01-01		&   011			& 012		&  regular update	\\
& 2021	& 11-01-01		&   012			& 013		&  regular update	\\
\hline	
{\bf BadPix}
& 1	& 00-01-01		&   012			& 012		&  values for launch \\
& 276	& 01-06-13		&   013			& 013		&  regular update \\
& 308	& 01-08-14		&   014			& 014		&  regular update \\
& 532	& 02-11-04		&			& 030		&  after cooling to -110$^\circ$C\\
& 536	& 02-11-13		&   015		 	& 		&  after cooling to -115$^\circ$C \\
& 543	& 02-11-27		&   030			&		&  after cooling to -110$^\circ$C  \\
& 1110	& 06-01-01		&   031			& 031		&  regular update including 2 hot regions in RGS1:CCD1 \\
& 1294	& 07-01-01		&			& 032           &  regular update \\
& 1417	& 07-04-09		&   032 		&		&  2 hot regions are  uploaded for orbit $>$ 1416\\
\hline
{\bf CoolPix}
& 1	& 00-01-01		&   001			& 001		&  values from launch \\
\hline
{\bf QuantumEff}
& 1	& 00-01-01		&   015			& 016		&  values from launch \\
\hline
{\bf EXAFS}
& 1	& 00-01-01		&   005			& 005		&  values from launch \\
\hline
{\bf EffAreaCorr}
& 1	& 00-01-01		&   008			& 008		&  values from launch \\
\hline
{\bf ReDist}
& 1	& 00-01-01		&   004			& 004		&  values from launch \\
\hline
{\bf LineSpreadFunc}
& 1	& 00-01-01		&   005			& 005		&  values from launch \\
\hline
{\bf CrossPSF}
& 1	& 00-01-01 		&   004 		& 004		&  values from launch		 \\
\hline
{\bf LinCoord}
& 1	& 00-01-01		&   008			& 008		&  support science simulator \\
\hline
{\bf TemplateBckGnd}
& 1	& 00-01-01		&   006			& 007		&  values at launch \\
& 1408	& 07-08-17		&			& 008		&  RGS2 single mode \\
\hline
{\bf SAACorr}
& 1	& 00-01-01		&   001			& 001		&  values from launch \\
\hline
{\bf HKParmint}
& 1	& 00-01-01		&   009			& 008		& values from launch \\
& 	& 00-09-04		&   010			& 		& \\
&	& 00-11-07		&   014			& 013		& \\
&	& 00-11-19		&   011			& 009		& \\
&	& 00-11-04		&			& 011		& \\
&	& 02-11-13		&   012			& 		& \\
&	& 02-11-14		&   013			& 012		& \\
\hline
{\bf XRT AreaEff}
& 17	& 00-01-13 		&   008			& 009    	&  initial effective area verified in orbit\\
\hline
{\bf XRT XPSF}
& 17	& 00-01-13 		&   014			& 014		&  values from launch \\
\hline
{\bf XMM BoreSight}
& 1	& 00-01-01 		&   023 		& 023 		&  values from launch \\
\hline
{\bf XMM MiscData}
& 1	& 00-01-01 		&   022			& 022		&  values from launch \\
\hline\noalign{\smallskip}
\end{supertabular}

\end{document}